\documentclass[english,journal=jctcce,manuscript=article,email=true,etalmode=truncate,maxauthors=0,doi=true]{achemso}

\usepackage{graphicx}
\graphicspath{{figures/}}
\usepackage{physics}
\usepackage{multirow}
\usepackage{xcolor}
\usepackage{caption}
\usepackage{subcaption}
\usepackage{xcolor}
\usepackage{bm}
\usepackage{eufrak} % for \mathfrak
\usepackage{yfonts} % for \textgoth
\usepackage[title,titletoc]{appendix}
\usepackage{diagbox}
\usepackage{multicol}
\usepackage{booktabs} % for tables obtained from pandas to_latex function
\usepackage{algorithm}
\usepackage{algpseudocode}
\usepackage{rotating}
\usepackage{threeparttable}

\usepackage[numbers,sort&compress,super]{natbib}
\usepackage{natmove}
\usepackage{hyperref} % hyperref must be placed before cleveref
\hypersetup{colorlinks=true, citecolor=blue, urlcolor=blue, linkcolor=blue}
\usepackage{cleveref}
  	\crefname{figure}{Figure}{Figures}
  	\crefname{table}{Table}{Tables}
  	\crefname{equation}{Eq.}{Eqs.}
  	\crefname{section}{Section}{Sections}
  	\crefname{subsection}{Section}{Sections}
  	\crefname{subsubsection}{Section}{Sections}
  	\crefname{algorithm}{Algorithm}{Algorithms}

\newcommand{\doi}[1]{\href{http://dx.doi.org/#1}{\nolinkurl{#1}}}

\newcommand\vartextvisiblespace[1][.5em]{%
  \makebox[#1]{%
    \kern.07em
    \vrule height.3ex
    \hrulefill
    \vrule height.3ex
    \kern.07em
  }% <-- don't forget this one!
}
\usepackage{todonotes}

\title{Revisiting Artifacts of Kohn-Sham Density Functionals for Biosimulation} %Title of paper

\author{Samuel A. Slattery}
\author{Jaden C. Yon}
\author{Edward F. Valeev}
\email{efv@vt.edu}
\affiliation{Department of Chemistry, Virginia Tech, Blacksburg, VA 24061}

\begin{document}

\begin{abstract}
We revisit the problem of unphysical charge density delocalization/fractionalization induced by the self-interaction error of common approximate Kohn-Sham Density Functional Theory functionals on simulation of small to medium-size proteins in vacuum. Aside from producing unphysical electron densities and total energies, the vanishing of the HOMO-LUMO gap associated with the unphysical charge delocalization leads to an unphysical low-energy spectrum and catastrophic failure of most popular solvers for the Kohn-Sham (KS) self-consistent field (SCF) problem. We apply a robust quasi-Newton SCF solver [{\em Phys. Chem. Chem. Phys.} {\bf 26}, 6557 (2024)] to obtain solutions for some of these difficult cases. The anatomy of the charge delocalization is revealed by the {\em natural deformation orbitals} obtained from the density matrix difference between the Hartree-Fock and KS solutions; the charge delocalization can occur not only between charged fragments (such as in zwitterionic polypeptides) but also may involve neutral fragments. The vanishing-gap phenomenon and troublesome SCF convergence are both attributed to the unphysical KS Fock operator eigenspectra of molecular fragments (e.g., amino acids or their side chains). Analysis of amino acid pairs suggests that the unphysical charge delocalization can be partially ameliorated by the use of {\em some} range-separated hybrid functionals, but not by semilocal or standard hybrid functionals. Last, we demonstrate that solutions without the unphysical charge delocalization can be located even for semilocal KS functionals highly prone to such defects, but such solutions have non-Aufbau character and are unstable with respect to mixing of the non-overlapping ``frontier'' orbitals.
Caution should be exercised when unexpectedly small (or vanishing) HOMO-LUMO gaps and atypical SCF convergence patterns (e.g., oscillatory) are observed in KS DFT simulations in any context (bio or otherwise).
\end{abstract}

\maketitle

\section{Introduction}
\label{section:introduction}

The unphysical behavior of all Kohn-Sham density functional approximations (DFAs) causes a number of artifacts, such as:
\begin{itemize}
    \item Unrealistically small HOMO-LUMO gaps\cite{VRG:tsuneda:2010:JCP,VRG:rudberg:2011:JCTC} and excitation energies,\cite{VRG:tozer:2003:JCP} especially for states with charge transfer\cite{VRG:dreuw:2003:JCP} and/or Rydberg\cite{VRG:casida:1998:JCP} characters (although some dispute these notions\cite{VRG:baerends:2013:PCCP}),
    \item charge fractionalization/delocalization, such as partial detachment of an electron in anions as the basis approaches completeness,\cite{VRG:jensen:2010:JCTC,VRG:peach:2015:JCTC}
    \item predicted reaction barriers too low,\cite{VRG:dobbs:1994:JPC,VRG:johnson:1994:CPL,VRG:zhang:1995:JPC,VRG:vydrov:2006:JCPa,VRG:janesko:2008:JCP}
    \item incorrect dissociation curves for both cations\cite{VRG:zhang:1998:JCP} and some neutrals,\cite{VRG:dutoi:2006:CPL}
    \item overbinding of opposite-charged moieties, \cite{VRG:otero-de-la-roza:2020:JPCA} and charge transfer complexes.\cite{VRG:ruiz:1996:JPC,VRG:isborn:2013:JPCB,VRG:otero-de-la-roza:2014:JCTC}
\end{itemize}
While these failures are well-recognized by the community, some artifactual behaviors of KS DFT in the context of biosimulations are less understood and invite more controversy.
For example, it was discovered that SCF convergence problems appeared for some biomolecules when using semilocal DFAs in vacuum.\cite{VRG:rubensson:2011:JCC,VRG:rudberg:2012:JPCM,VRG:antony:2012:JCC,VRG:kulik:2012:JPCB}
Subsequent work concluded that no fundamental problem exists for semilocal DFAs when applied to such systems, and that the problem was rather with the way the system was ``prepared".\cite{VRG:lever:2013:JPCM,VRG:cole:2016:JPCM}
Ultimately this came down to two major factors: missing solvent effects and geometry relaxation.
Thus, commonly proposed solutions for the problem include: (1) use of a solvent model (particularly implicit\cite{VRG:cole:2016:JPCM,VRG:zuehlsdorff:2016:JCTC,VRG:ren:2022:JCP} or even point charges\cite{VRG:rudberg:2012:JPCM}), (2) application of geometry relaxation,\cite{VRG:lever:2013:JPCM,VRG:cole:2016:JPCM} and (3) change to range-separated functionals\cite{VRG:kulik:2012:JPCB,VRG:sepunaru:2015:JACS,VRG:sharley:2016:} (e.g. $\omega$B97).
We agree that the problem is in the functional, but note here that even range-separated functionals do not systematically solve these problems for all cases.\cite{VRG:vydrov:2006:JCPb}
%For example, the range-separation parameter often is tuned for specific systems.\sas{cite}
The idea that the fundamental problem is electrostatic in nature, rather than a result of more essential problems with the approximate functionals has persisted.\cite{VRG:li:2015:PCCP,VRG:zuehlsdorff:2016:JCTC}

Many researchers have realized that some of these problems can have striking consequences for the use of KS DFT for systems containing charged moieties. For example, Jensen\cite{VRG:jensen:2010:JCTC} recognized that the artificially high HOMO energy of the negatively charged deprotonated carboxylic acid group in a peptide zwitterion in vacuum could result in fractional charge transfer.
While it has long been known that approximate KS DFT struggles with anions,\cite{VRG:shore:1977:PRB,VRG:kschwarz:1978:JPBAMP,VRG:schwarz:1978:CPL,VRG:rosch:1997:JCP,VRG:peach:2010:JPCL,VRG:jensen:2010:JCTC,VRG:lee:2010:JPCL,VRG:kim:2011:JCP,VRG:rudberg:2012:JPCM} the work of Jensen indicated the possibility of transfer of electron density within a single molecule.
This possibility relates to the idea of ``delocalization error"\cite{VRG:mori-sanchez:2006:JCP,VRG:mori-sanchez:2008:PRL,VRG:cohen:2008:S} where the density is too spread out due to the nonlinearity of the KS DFT total energy with respect to fractional electron number.\cite{VRG:perdew:1982:PRL,VRG:zhang:1998:JCP,VRG:mori-sanchez:2006:JCP,VRG:peach:2015:JCTC,VRG:hait:2018:JPCL}
Clearly, this can result in incorrect electrostatic descriptions,\cite{VRG:jakobsen:2013:JCTC,VRG:lever:2013:JPCM} but it can also cause the aforementioned SCF convergence problems.\cite{VRG:rubensson:2011:JCC,VRG:lever:2013:JPCM}
One proposed solution to some of these issues involves evaluating KS DFT energies using HF densities,\cite{VRG:lee:2010:JPCL,VRG:kim:2011:JCP,VRG:sim:2022:JACS} although the lack of self-consistency would then be a substantial concern.
%However, as has been pointed out,\sas{cite} this is not optimal since the energy and density are obtained via different methods.

In this work, we revisit the convergence problems of SCF for local and semilocal DFAs applied to medium-sized biomolecules in vacuum.
In \cref{section:pdb_systems} we examine solutions for a set of 17 polypeptides considered in Ref. \citenum{VRG:rudberg:2012:JPCM} using a robust local-convergence SCF solver (QUOTR\cite{VRG:slattery:2024:PCCP}), thus allowing us to characterize the ``true" (energy minimized) solutions even for cases where no solutions could be found in Ref. \citenum{VRG:rudberg:2012:JPCM}; the SCF convergence issues are correlated to the unphysically small (or even vanishing) HOMO-LUMO gap. Although the incorrect density predicted by approximate functionals has been investigated for zwitterions,\cite{VRG:jakobsen:2013:JCTC,VRG:ren:2022:JCP} the orbital structure (at least for the semilocal functionals) was not examined in detail. \cref{section:pdb_systems} includes details of orbital-based analysis of KS$\to$HF deformation densities revealing in pinpoint detail the anatomy of unphysical KS fractionalization/delocalization of charges in such systems. The use of deformation density to decompose a change in density into orbitals contributions has precedent in the context of fragment to complex interaction,\cite{VRG:pakiari:2008:IJQC} but it seems to be unknown for comparing different quantum chemistry methods.
Additionally, we show that by using a local solver we can obtain SCF solutions with properly-localized charges in some cases, which are actually non-Aufbau (and non-energy minimizing).
We follow this in \cref{section:amino_acids} by scanning through several popular functionals for all 20 naturally occurring amino acids to reveal which combinations of amino acids with which DFAs can cause such issues.
In particular, we note that range-separation, while almost always helpful, does not eliminate the possibility of unphysical charge delocalization and problematic SCF convergence.

\section{Technical Details}
\label{section:technical_details}

HF and KS DFT computations on polypeptides in \cref{section:pdb_systems} were performed with a developmental version of the Massively Parallel Quantum Chemistry (MPQC) version 4 program package,\cite{VRG:peng:2020:JCP} using the recently developed QUOTR SCF solver.\cite{VRG:slattery:2024:PCCP}
The maximum L-BFGS history size was set to 15 (parameter $m$), and the initial guess was the unperturbed version of the extended-H{\"u}ckel-like guess used previously, except for 1RVS which used the perturbed guess.
The regularizer threshold ($t_r$) was lowered to 0.15, and the history was also reset whenever the RMS gradient crossed $1 \times 10^{-6}$ (either crossing below or coming back up again). 
However, the calculations on the glycine systems in \cref{section:gly_separated} used the same parameters as in ref. \citenum{VRG:slattery:2024:PCCP}.
One final difference in the solver is that the history data was trimmed whenever the lowest eigenvalue of the matrix $\mathbf{V}^{\mathrm{T}}\mathbf{V}$ was below $-1\times 10^{-15}$ until that condition was no longer true. %, all other parameters are the same as in ref. \citenum{VRG:slattery:2024:PCCP}.\sas{check this}
The KS DFT implementation in MPQC employs GauXC\cite{VRG:petrone:2018:EPJB} (which calls LibXC\cite{VRG:lehtola:2018:S}) for calculation of the exchange-correlation potentials and energies.
The integration grid for evaluation of these quantities for the PDB systems was the ``superfine'' grid (250 radial Mura-Knowles \cite{VRG:mura:1996:JCP} points, 974 angular Lebedev-Laikov \cite{VRG:lebedev:1999:DM} for all atoms except hydrogen, which has 175 radial points.).
Density functionals used include LDA (SVWN5),\cite{VRG:dirac:1930:MPCPS,VRG:vosko:1980:CJP} BLYP,\cite{VRG:miehlich:1989:CPL} PBE,\cite{VRG:perdew:1996:PRL} B3LYP,\cite{VRG:stephens:1994:JPC} PBE0,\cite{VRG:adamo:1999:JCP,VRG:ernzerhof:1999:JCP} for the PDB systems and additionally for the peptide pair screening in \cref{section:amino_acids_screening}: revPBE,\cite{VRG:zhang:1998:PRL} MN15,\cite{VRG:yu:2016:CS} TPSS,\cite{VRG:tao:2003:PRL} SCAN,\cite{VRG:sun:2015:PRL} revSCAN,\cite{VRG:mezei:2018:JCTC} CAM-B3LYP,\cite{VRG:yanai:2004:CPL} $\omega$B97,\cite{VRG:chai:2008:JCP} $\omega$PBE.\cite{VRG:henderson:2008:JCP,VRG:weintraub:2009:JCTC}
To match the calculations performed in ref. \citenum{VRG:rudberg:2012:JPCM} we used the 6-31G** basis.\cite{VRG:ditchfield:1971:JCP,VRG:hehre:1972:JCP,VRG:hariharan:1973:TCA,VRG:francl:1982:JCP,VRG:gordon:1982:JACS}
Density fitting was performed with the def2-universal-J basis. \cite{VRG:weigend:2006:PCCP}
Geometries for all systems (except the individual amino acids used in \cref{section:amino_acids}) were obtained from the Protein Data Bank (PDB).\cite{VRG:berman:2000:NAR}

The electron densities of the converged KS DFT solutions were analyzed by comparing them with the corresponding HF electron densities.
This was done by finding the eigenvalues and eigenvectors of the difference of the density matrices (in orthonormal basis).
\begin{align}
\mathbf{\gamma}^{\mathrm{diff}} = \mathbf{\gamma}^{\mathrm{HF}} - \mathbf{\gamma}^{\mathrm{KS}}
\end{align}
These eigenvectors were then transformed back to the AO basis; these are the HF-KS \textit{natural deformation orbitals} (NDOs).
The eigenvalues associated with each NDO will be referred to as natural deformation charges $q_\mathrm{nd}$. Plotting NDOs with negative $q_\mathrm{nd}$ identifies the regions where KS has gained density (relative to HF), and vice versa.
Although the use of a post-HF reference density would be preferred to include correlation effects (e.g., MP2), for the cases of qualitative failure of KS DFT the HF-KS and MP2-KS natural deformation orbitals should be qualitatively similar.

Amino acid calculations in \cref{section:amino_acids_screening} used Psi4\cite{VRG:smith:2020:JCP} for both geometry optimization (B3LYP/6-31G*\cite{VRG:ditchfield:1971:JCP,VRG:hehre:1972:JCP,VRG:hariharan:1973:TCA,VRG:francl:1982:JCP,VRG:gordon:1982:JACS}) and single point KS DFT energy evaluation (along with eigenspectrum), using default parameters, including density fitting.
The side chains of the amino acids were uncharged in all cases, except for arginine, which had a +1 charge.

\section{Results}
\label{section:results}

\subsection{Small polypeptides}
\label{section:pdb_systems}

Our analysis starts with the set of 17 small polypeptides used by Rudberg to illustrate SCF convergence failures.\cite{VRG:rudberg:2012:JPCM} For 12 of these systems the standard diagonalization-based Roothaan-Hall (RH) SCF solver could not locate a solution for at least one functional; the chemical structures for these 12 ``difficult'' systems are presented in the SI.
Using our QUOTR solver we managed to obtain converged SCF solution for all system/DFA combinations considered by Rudberg (BHandHLYP was excluded since it posed no convergence issues). For all RH-converged SCF solutions in Ref. \citenum{VRG:rudberg:2012:JPCM} QUOTR confirmed the HOMO-LUMO gaps to within 0.04 eV (the largest deviation was observed for the 1XT7 system with the LDA functional). For the cases where the RH solver could not locate a solution the QUOTR solver located solutions with vanishing HOMO-LUMO gap. These vanishing-gap QUOTR solutions were then analyzed by the HF-KS deformation density analysis (DDA) described in \cref{section:technical_details}.

\begin{sidewaystable}
% Data Source: <Jaden>
    \begin{tabular}{lcccccccccccc}
    \toprule
    \toprule
    {} & \multicolumn{2}{c}{HF} & \multicolumn{2}{c}{PBE0} & \multicolumn{2}{c}{B3LYP} & \multicolumn{2}{c}{BLYP} & \multicolumn{2}{c}{PBE} & \multicolumn{2}{c}{LDA}\\ \cmidrule(lr){2-3} \cmidrule(lr){4-5} \cmidrule(lr){6-7} \cmidrule(lr){8-9} \cmidrule(lr){10-11} \cmidrule(lr){12-13}
     PDB ID & RH & QUOTR & RH & QUOTR & RH & QUOTR & RH & QUOTR & RH & QUOTR & RH & QUOTR \\
    \midrule
    2P7R & 12.03 & 12.04 & 4.65 & 4.67 & 4.16 & 4.18 & 2.12 & 2.14 & 2.10 & 2.11 & 2.12 & 2.14 \\
    1BFZ & 11.96 & 11.96 & 6.18 & 6.18 & 5.77 & 5.77 & 3.97 & 3.98 & 3.93 & 3.93 & 3.79 & 3.79 \\
    2IGZ & 11.81 & 11.81 & 5.70 & 5.70 & 5.27 & 5.27 & 3.27 & 3.28 & 3.23 & 3.24 & 3.12 & 3.13 \\
    1D1E & 10.14 & 10.12 & 3.96 & 3.96 & 3.47 & 3.47 & 1.56 & 1.57 & 1.58 & 1.59 & 1.54 & 1.55 \\
    1SP7 & 9.13 & 9.11 & 1.65 & 1.64 & 0.87 & 0.87 & \textemdash & $<0.01$ & \textemdash & $<0.01$ & \textemdash & $<0.01$ \\
    1N9U & 9.12 & 9.12 & 1.12 & 1.14 & 0.57 & 0.59 & \textemdash & $<0.01$ & \textemdash & $<0.01$ & \textemdash & $<0.01$ \\
    1MZI & 8.77 & 8.74 & 1.29 & 1.29 & 0.54 & 0.54 & \textemdash & $<0.01$ & \textemdash & $<0.01$ & \textemdash & $<0.01$ \\
    1XT7 & 8.51 & 8.48 & 3.32 & 3.29 & 2.65 & 2.63 & 1.02 & 1.00 & 1.24 & 1.23 & 1.30 & 1.34 \\
    1PLW & 7.25 & 7.24 & 0.36 & 0.36 & 0.29 & 0.28 & \textemdash & $<0.01$ & \textemdash & $<0.01$ & \textemdash & 0.01 \\
    1FUL & 6.95 & 6.93 & 0.20 & 0.20 & 0.16 & 0.16 & \textemdash & $-0.01$ & \textemdash & $-0.01$ & \textemdash & $-0.01$ \\
    1EDW & 6.89 & 6.90 & 0.26 & 0.26 & 0.21 & 0.21 & \textemdash & $<0.01$ & \textemdash & $<0.01$ & \textemdash & $<0.01$ \\
    1EVC & 5.82 & 5.84 & 0.30 & 0.30 & 0.24 & 0.24 & \textemdash & $<0.01$ & \textemdash & $<0.01$ & \textemdash & $<0.01$ \\
    1RVS & 5.60 & 5.59 & \textemdash & 0.10 & \textemdash & 0.08 & \textemdash & $<0.01$ & \textemdash & $<0.01$ & \textemdash & $<0.01$ \\
    2FR9 & 5.48 & 5.50 & 0.26 & 0.26 & 0.21 & 0.21 & \textemdash & $<0.01$ & \textemdash & $<0.01$ & \textemdash & $<0.01$ \\
    2JSI & 5.26 & 5.27 & 0.24 & 0.24 & 0.19 & 0.19 & \textemdash & $<0.01$ & \textemdash & $<0.01$ & \textemdash & $<0.01$ \\
    1LVZ & 5.05 & 5.03 & 0.31 & 0.31 & 0.25 & 0.25 & \textemdash & $<0.01$ & \textemdash & $<0.01$ & \textemdash & $<0.01$ \\
    1FDF & 3.64 & 3.62 & 0.13 & 0.14 & 0.11 & 0.11 & \textemdash & $<0.01$ & \textemdash & $<0.01$ & \textemdash & $<0.01$ \\
    \bottomrule
    \bottomrule
    \end{tabular}
    \caption{HOMO-LUMO gaps (eV) for 17 small polypeptides from Ref. \citenum{VRG:rudberg:2012:JPCM}. The standard (RH) SCF solver values are from \citenum{VRG:rudberg:2012:JPCM}, the quasi-Newton (QUOTR) SCF solver\cite{VRG:slattery:2024:PCCP} values are from this work. For all cases where the standard solver failed to converge to a solution, the quasi-Newton solver located a solution has a vanishing HOMO-LUMO gap.}
    \label{table:pdb_systems}
\end{sidewaystable}

\subsubsection{HF-LDA deformation density analysis summary}
\label{section:overview_analysis}

Although Hartree-Fock is known to artificially localize charges,\cite{VRG:mori-sanchez:2008:PRL} it provides qualitatively correct electron distributions for all these systems; specifically, the HF charges agree with the expected integral values of formal charges on charged functional groups like -CO$_2^-$. In all cases HF predicts large positive HOMO-LUMO gaps. In contrast, KS DFT can ``delocalize'' charges across large distances, or alternatively it can lead to fractional charges on charged functional group; such states have vanishing HOMO-LUMO gap.
The HF-KS DDA reveals where the fractional charges are located in the KS solution.
The summary of HF-LDA DDA performed for the 12 systems with vanishing LDA gap (\cref{table:pdb_analysis_overview}) reports which and how many functional groups donate ($N_\mathrm{d}$) and accept ($N_\mathrm{a}$) electron density in LDA relative to HF.
Usually, but not always, the donor and acceptor sites correspond to the location of the Hartree-Fock HOMO and LUMO, or other states near the Fermi level.
See the SI for images comparing the Hartree-Fock HOMO and LUMO for each system with the natural deformation orbitals used in our analysis.
In all cases shown in \cref{table:pdb_analysis_overview}, the electron donors are formally negatively charged moieties (mostly CO$_2^-$).
The electron acceptors are often positively charged (mostly NH$_3^+$), however, a few cases involve neutral functional groups as acceptors, most notably 2FR9, 1EDW, 1EVC, and 2JSI.
Sometimes the natural deformation orbitals are not so clear to interpret.
The column labeled ``backbone?" is used to indicate when there are some density change contributions that are hard to classify which involve other parts of the system.
In most cases, these contributions include carbonyl groups in the peptide bonds.
    
\cref{table:pdb_analysis_overview} also characterizes the HF-LDA NDOs with nonnegligible $q_\mathrm{nd}$, such orbitals will be referred to as frontier NDOs (FNDOs). For all systems studied, the smallest $q_\mathrm{nd}$ was simply the negative of the largest one, thus we only display the largest $|q_\mathrm{nd}|$ for each system. In many cases there is a single pair of FNDOs, which typically corresponds to the case of electrons transferred between single functional groups. But in some systems there are multiple FNDOs (e.g. 1FUL); in such cases we only display the largest $|q_\mathrm{nd}|$.
We only record donor/acceptor groups in \cref{table:pdb_analysis_overview} contributing to the FNDOs with $|q_\mathrm{nd}| \geq 0.2$.
Only three systems had multiple $|q_\mathrm{nd}|$ over this threshold: 1FUL, 2F9R, and 1FDF.
Images of all natural deformation orbitals used in the analysis are provided in the SI.

\begin{table}
% Data Source: <Jaden>
    \begin{tabular}{l|ccccc|c}
    \toprule
    \toprule
    PDB ID & $|q_\mathrm{nd}|$ & $N_\mathrm{d}$ & donor types & $N_\mathrm{a}$ & acceptor types & backbone? \\ 
    \hline
    1SP7 & 0.211 & 1 & $-$CO$_2^-$ & 1 & $-$NH$_3^+$ & minimal \\
    1N9U & 0.282 & 2 & $-$CO$_2^-$ & 1 & $-$NHC(NH$_2)_2^+$ & yes \\
    1MZI & 0.284 & 2 & $-$CO$_2^-$ & 1 & $-$NH$_3^+$ & minimal \\
    1PLW & 0.379 & 1 & $-$CO$_2^-$ & 1 & $-$NH$_3^+$ & some \\
    1FUL & 0.411 & 2 & $-$CO$_2^-$ & 2 & $-$NH$_3^+$,$-$NHC(NH$_2)_2^+$ & some;S$_2$ \\
    1EDW & 0.414 & 2 & $-$CO$_2^-$ & 1 & phenyl near $-$NH$_3^+$ & some \\
    1EVC & 0.447 & 1 & $-$CO$^-$ & 1 & phenyl & yes \\
    1RVS & 0.439 & 1 & $-$CO$_2^-$ & 1 & $-$NH$_3^+$ and phenyl & minimal \\
    2FR9 & 0.566 & 2 & $-$CO$^-$,$-$NH$^-$ & 2 & phenyl,amide & yes;S$_2$  \\
    2JSI & 0.468 & 1 & $-$NH$^-$ & 1 & phenol & yes \\
    1LVZ & 0.687 & 3 & $-$CO$_2^-$ & 2 & $-$NH$_3^+$,$-$NHC(NH$_2)_2^+$ & some \\
    1FDF & 0.564 & 2 & $-$CO$^-$,$-$PRO$^-$ & 3 & $-$NH$_3^+$,$-$NHC(NH$_2)_2^+$,amide & yes \\
    \bottomrule
    \bottomrule
    \end{tabular}
    \begin{tablenotes}
        \item $^a$ $N_\mathrm{d}$ is ``number of donors"; $N_\mathrm{a}$ is ``number of acceptors". See text for further details.
        \item $^b$ $-$NHC(NH$_2)_2^+$ is protinated guanidine; $-$PRO is proline.
    \end{tablenotes}
    \caption{HF-LDA natural deformation density analysis$^a$ for the 12 systems with SCF convergence difficulties from Ref. \citenum{VRG:rudberg:2012:JPCM}, indicating donor group types and acceptor group types.$^b$}
    \label{table:pdb_analysis_overview}
\end{table}

We next examine three different cases that provide insight into the possible scenarios.

\subsubsection{Case 1FUL: 2 donors, 2 acceptors}
\label{section:case1_simple}

\begin{figure}
% Data Source: <Jaden>
    \centering
    \includegraphics[width=6.0in]{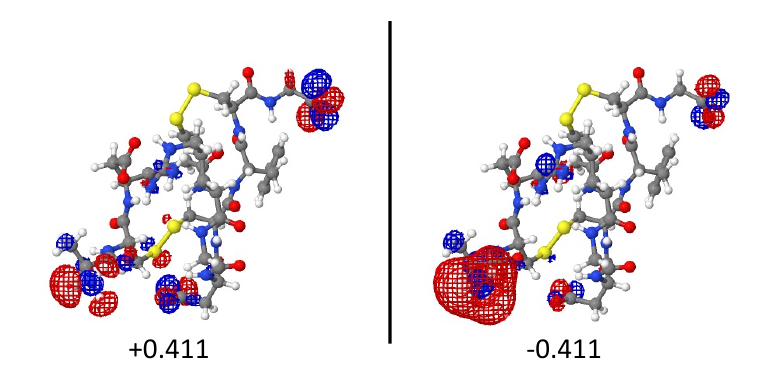}
    \caption{The HF-LDA NDOs of 1FUL with largest-magnitude deformation charges. This case illustrates charge delocalization over multiple donor and multiple acceptor functional groups.}
    \label{fig:1FUL_orb1}
\end{figure}

First, we consider a zwitterion that exemplifies the classic case of charge separation, where the delocalization error is known to be a problem.\cite{VRG:jakobsen:2013:JCTC,VRG:ren:2022:JCP}
In \cref{fig:1FUL_orb1} we display the natural deformation orbitals with the largest deformation charges, $q_\mathrm{nd} = \pm 0.411$.
In this case there are actually at least two regions that appear to be accepting electrons: the NH$_3^+$ group and the protonated guanidine group of arginine.
Also, there are two deprotonated carboxylic acid groups that donate electrons.
Although this case is very similar to the classic zwitterions, the analysis is made more complicated because there is density transfer occurring with the peptide backbone and at least one of the disulfide bridges.

\subsubsection{Case 1EVC: neutral acceptor}
\label{section:case2_neutral_receiver}

\begin{figure}
% Data Source: <Jaden>
    \centering
    \includegraphics[width=7.0in]{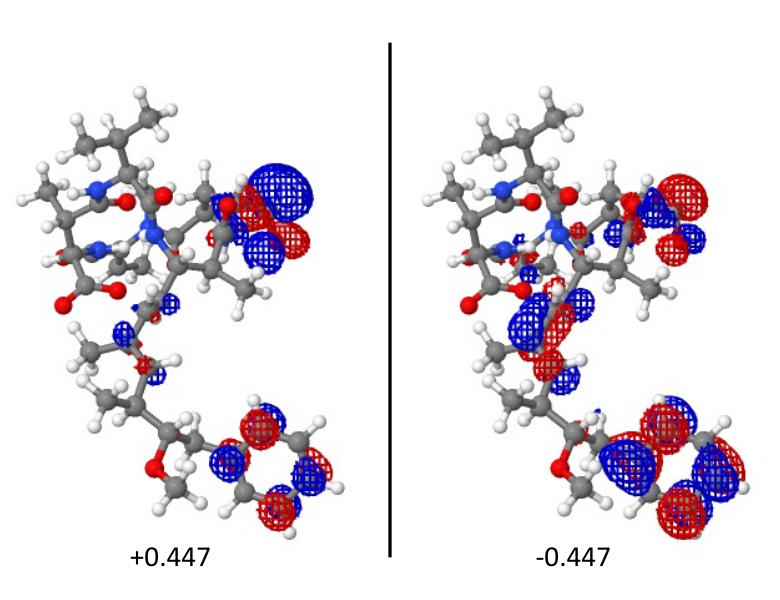}
    \caption{The HF-LDA natural deformation orbitals of 1EVC with largest-magnitude deformation charges.}
    \label{fig:1EVC_orb1}
\end{figure}

The problems with zwitterion convergence using semilocal DFAs are relatively well known.
Here we show that it is possible to have the same problem without an explicitly positive group receiving the electron density.
In \cref{fig:1EVC_orb1} the natural deformation orbitals of largest magnitude are displayed for 1EVC.

Closer examination of the structure reveals that electron density is being donated from a CO$^-$ (not CO$^-_2$) group!
Clearly this system has an invalid structure generated to fit the experimental NMR data.
For such unphysical structures it is possible to have an extremely unstable anion whose density will be transferred to a neutral group. However, such exceptions are rare since this is one of the few examples of anion-to-neutral charge transfer in the current test set.

\subsubsection{Case 1RVS: non-Aufbau solution}
\label{section:case3_nonaufbau}

\begin{figure}
% Data Source: <Jaden>
    \centering
    \includegraphics[width=5.25in]{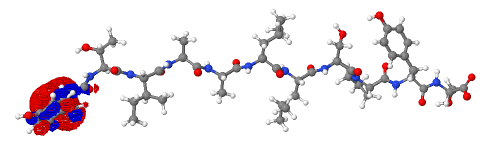}
    \includegraphics[width=5.25in]{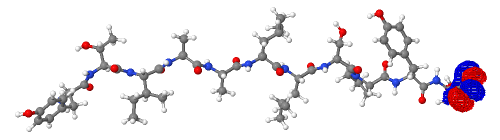}
    \caption{For the 1RVS zwitterion it was possible to locate a non-Aufbau B3LYP SCF soluton that does not suffer from charge delocalization and qualitatively matches the charge distribution of the exact ground state. The non-Aufbau HOMO (bottom) and LUMO (top) are localized on the CO$_2^-$ and NH$_3^+$ termini, as expected.}
    \label{fig:1RVS_homo-lumo}
\end{figure}

Although the near-zero gap displayed for 1FUL in \cref{section:case1_simple} is typical for zwitterions in vacuum, we have found that for a different zwitterion our local SCF solver is able to converge to the ``physically correct" charge-separated solution.
In \cref{fig:1RVS_homo-lumo} we display the HOMO and LUMO for 1RVS when the QUOTR solver is given the usual extended-H{\"u}ckel-like initial guess for the orbitals, but without perturbation.\cite{VRG:slattery:2024:PCCP}
We see that there is no unphysical mixing of the occupied orbital on the CO$_2^-$ with the unoccupied orbital on the NH$_3^+$.
For this solution the HOMO has an energy of -0.037 eV while the LUMO has an energy of -0.130 eV, giving a nonzero negative HOMO-LUMO gap.
Thus, this is a non-Aufbau state!
Now, this solution is not the best solution in a variational sense, because the total energy could be lowered by mixing the HOMO with the LUMO (note Janak's theorem\cite{VRG:janak:1978:PRB}).
However, it does avoid the major contribution to the incorrect charge delocalization.
We conclude that we have been able to find a local stationary point for 1RVS (a zwitterion system) using B3LYP that has qualitatively the physically correct HOMO and LUMO, but such solution is not the lowest energy state.
When the initial guess is perturbed, QUOTR is able to converge to the nearly zero-gap solution.
The energy of this solution is 0.012 $E_{\mathrm{h}}$ below the physically-reasonable non-Aufbau solution.

\subsection{Amino Acids}
\label{section:amino_acids}

The simplest analogous situation where the HOMO could be above the LUMO is in a system composed of two amino acids (in vacuum) separated by a large distance so that they are essentially non-interacting.
This simple model is explored in \cref{section:gly_separated} where we show that the non-Aufbau solution is obtainable using the QUOTR SCF solver.
By perturbing the guess orbitals (importantly mixing orbitals on the separated fragments), we can also obtain the near zero-gap solution.
The underlying principle thus indicates that this will occur whenever the HOMO on one fragment is above the LUMO of the other fragment.
Therefore, in \cref{section:amino_acids_screening} we calculate HOMO and LUMO energies for a variety of popular functionals, to see which combinations are possibly problematic.

\subsubsection{Separated glycine zwitterion}
\label{section:gly_separated}

We demonstrate the utility of QUOTR by finding a solution for a system that is impossible for a diagonalization-based solver: a non-Aufbau filled system.
For this analysis we use KS-DFT with the local density approximation (LDA) to the exchange-correlation functional.

\begin{figure}
% Data Source: /Users/sas/research_data/raw/2024_03_26/run_gly-sep_quotr_6-31gs
% images: /Users/sas/Documents/my_papers/peptides/images/structures_jmol_figure.png
    \centering
    \includegraphics[width=3.25in]{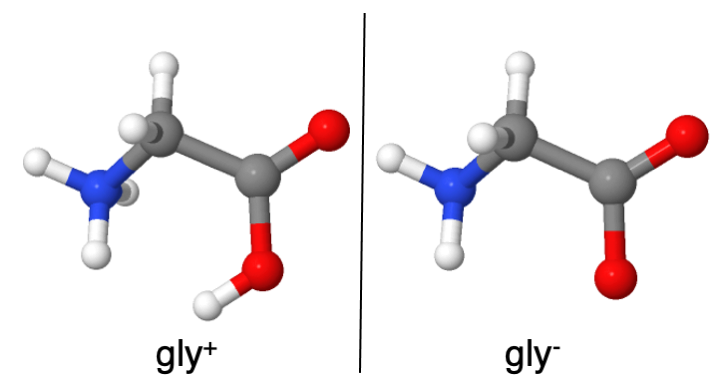}
    \caption{Glycine in protonated (overall positive charge, gly$^+$) and deprotonated (overall negative charge, gly$^-$) states.}
    \label{fig:gly_xyz}
\end{figure}

The HOMO and LUMO for glycine in two protonation states (labled gly$^{-}$ and gly$^{+}$, see \cref{fig:gly_xyz}) were calculated separately.
These could represent the two ends of a peptide chain in zwitterion form.
Then a supersystem was constructed with these two fragments together, separated by approximately 200 angstrom.
Thus, the two fragments should be physically isolated from each other. 

From the data in \cref{table:glycine_analysis} we see that a non-Aufbau solution is expected: although each charged fragment has a sizable positive HOMO-LUMO gap, the HOMO of the anion is much higher than the LUMO of the cation.
When these two fragments are calculated together, the initial guess for the MOs is of utmost importance for any direct minimization solver.
In the column labeled ``unperturbed" the initial guess is our standard extended-H\"{u}ckel guess. 
Due to the large separation of about 200 angstroms, there is no overlap between the AOs of the different fragments.
Thus, the guess MOs are disjoint: each orbital is either associated with gly$^{-}$ or gly$^{+}$.
The gradient for mixing the MOs on different fragments should therefore be zero.
Minimization should then result in nearly the same orbitals on each fragment as in the isolated calculations, implying a solution with the HOMO higher than the LUMO!
In contrast, the column labeled ``perturbed" takes the extended-H\"{u}ckel guess and applies a small, random unitary matrix that allows mixing between all orbitals, regardless of which fragment they belong to.
Thus, there are likely some orbitals in the initial guess that span both fragments.
This is physically incorrect, but LDA is actually able to find a lower total energy for the supersystem with some MOs spanning both fragments.
The HOMO is then lowered and the LUMO raised, until the gap is essentially zero.

\begin{table}
% Data Sources: /Users/sas/research_data/raw/2024_03_26/run_gly-sep_quotr_6-31gs
    \centering
    \begin{tabular}{rccccc}
    
    \toprule
    \toprule
    {} & gly$^{-}$ & gly$^{+}$ & {} & \multicolumn{2}{c}{gly$^{-} \dots$gly$^{+}$} \\ \cmidrule{5-6}
     & & & {} & unperturbed & perturbed \\
    \midrule
    LUMO: & 5.72 & -6.79 & {} & -6.72 & -3.31 \\
    HOMO: & 0.64 & -12.59 & {} & 0.57 & -3.29 \\
    HOMO-LUMO gap: & 5.08 & 5.80 & {} & -7.29 & -0.02 \\
    \bottomrule
    \bottomrule
    \end{tabular}
    \caption{Frontier orbital energetics (eV) of LDA/6-31G** SCF solutions for isolated deprotonated/protonated glycine molecules and their noninteracting supersystem. Quasi-Newton solver locates a non-Aufbau supersystem solution, unless the initial (atomic density-based) guess is perturbed to mix the orbitals of the monomers. Perturbed guess leads to the solution with a vanishing HOMO-LUMO gap, and an unphysical charge delocalization.}
    \label{table:glycine_analysis}
\end{table}

This striking example is due to the self-interaction error in the LDA functional, and it demonstrates LDA's tendency to prefer fractional electron systems.
In this case the fractional electron on one of the fragments is caused by having an MO with substantial density on the other fragment too.

\subsubsection{Screening natural amino acid pairs}
\label{section:amino_acids_screening}

To systematically investigate when the problem of vanishing HOMO-LUMO gap is likely to appear, we do not need to run the same calculations as in the previous section for each pair of amino acids. 
The principle is clear: when an isolated fragment has a HOMO above the LUMO of a separate fragment, a non-Aufbau filled solution is possible.
The idea that having the LUMO on one fragment above the HOMO on another other fragment can cause excessive electron transfer and binding strength has been examined before in the context of radical-molecules complexes,\cite{VRG:johnson:2013:JPCA} and charge transfer complexes.\cite{VRG:sini:2011:JCTC}
We performed a series of tests on different functionals for all 20 naturally occurring amino acids in three different protonation states: neutral, anionic, and cationic (except for Arginine, where the charges are shifted to 0, +1, and +2).
\Cref{table:non-aufbau_systems} displays the number of predicted non-Aufbau pairs of amino acids, grouped by charge state, for a representative set of local, gradient-corrected (GGA), meta-GGA, and hybrid DFAs (both standard and range-separated).
    
\begin{table}
% Data Source: /Users/sas/research_data/processed/ionization_potential/amino_acids 
% Note: only CAM-B3LYP and wB97 are NOT at the above location
    \begin{tabular}{cr|ccccrr|cccccr|cccc}
    \toprule
    \toprule
    \multicolumn{18}{c}{Semilocal} \vspace{0.1 cm}\\
    \multicolumn{2}{c}{LDA} & \multicolumn{4}{c}{$Q_\mathrm{a}$} & \multicolumn{2}{c}{BLYP} & \multicolumn{4}{c}{$Q_\mathrm{a}$} & \multicolumn{2}{c}{revPBE} & \multicolumn{4}{c}{$Q_\mathrm{a}$} \\ 
    \multicolumn{2}{c}{} & -1 & 0 & +1 & +2 & \multicolumn{2}{c}{} & -1 & 0 & +1 & +2 & \multicolumn{2}{c}{} & -1 & 0 & +1 & +2 \\ \cmidrule(lr){3-6} \cmidrule(lr){9-12} \cmidrule(lr){15-18}
    \multirow{4}{0.2 cm}{$Q_\mathrm{d}$} & -1 & 0 & 379 & 380 & 19 & \multirow{4}{0.2 cm}{$Q_\mathrm{d}$} & -1 & 0 & 380 & 380 & 19 & \multirow{4}{0.2 cm}{$Q_\mathrm{d}$} & -1 & 0 & 380 & 380 & 19 \\ 
    {} & 0 & 0 & 0 & 306 & 20 & {} & 0 & 0 & 0 & 286 & 20 & {} & 0 & 0 & 0 & 273 & 20 \\ 
    {} & +1 & 0 & 0 & 0 & 3 & {} & +1 & 0 & 0 & 0 & 4 & {} & +1 & 0 & 0 & 0 & 2 \\ 
    {} & +2 & 0 & 0 & 0 & 0 & {} & +2 & 0 & 0 & 0 & 0 & {} & +2 & 0 & 0 & 0 & 0 \\ 
    \multicolumn{18}{c}{} \vspace{0.05 cm}\\
    \hline

    \multicolumn{18}{c}{meta-GGA} \vspace{0.1 cm}\\
    \multicolumn{2}{c}{TPSS} & \multicolumn{4}{c}{$Q_\mathrm{a}$} & \multicolumn{2}{c}{SCAN} & \multicolumn{4}{c}{$Q_\mathrm{a}$} & \multicolumn{2}{c}{revSCAN} & \multicolumn{4}{c}{$Q_\mathrm{a}$} \\ 
    \multicolumn{2}{c}{} & -1 & 0 & +1 & +2 & \multicolumn{2}{c}{} & -1 & 0 & +1 & +2 & \multicolumn{2}{c}{} & -1 & 0 & +1 & +2 \\ \cmidrule(lr){3-6} \cmidrule(lr){9-12} \cmidrule(lr){15-18}
    \multirow{4}{0.2 cm}{$Q_\mathrm{d}$} & -1 & 0 & 368 & 380 & 19 & \multirow{4}{0.2 cm}{$Q_\mathrm{d}$} & -1 & 0 & 275 & 380 & 19 & \multirow{4}{0.2 cm}{$Q_\mathrm{d}$} & -1 & 0 & 265 & 380 & 19 \\ 
    {} & 0 & 0 & 0 & 176 & 20 & {} & 0 & 0 & 0 & 74 & 20 & {} & 0 & 0 & 0 & 63 & 20 \\ 
    {} & +1 & 0 & 0 & 0 & 1 & {} & +1 & 0 & 0 & 0 & 0 & {} & +1 & 0 & 0 & 0 & 0 \\ 
    {} & +2 & 0 & 0 & 0 & 0 & {} & +2 & 0 & 0 & 0 & 0 & {} & +2 & 0 & 0 & 0 & 0 \\ 
    \multicolumn{18}{c}{} \vspace{0.05 cm}\\
    \hline

    \multicolumn{18}{c}{Hybrid} \vspace{0.1 cm}\\
    \multicolumn{2}{c}{MN15$^a$} & \multicolumn{4}{c}{$Q_\mathrm{a}$} & \multicolumn{2}{c}{B3LYP} & \multicolumn{4}{c}{$Q_\mathrm{a}$} & \multicolumn{2}{c}{PBE0} & \multicolumn{4}{c}{$Q_\mathrm{a}$} \\ 
    \multicolumn{2}{c}{} & -1 & 0 & +1 & +2 & \multicolumn{2}{c}{} & -1 & 0 & +1 & +2 & \multicolumn{2}{c}{} & -1 & 0 & +1 & +2 \\ \cmidrule(lr){3-6} \cmidrule(lr){9-12} \cmidrule(lr){15-18}
    \multirow{4}{0.2 cm}{$Q_\mathrm{d}$} & -1 & 0 & 0 & 380 & 19 & \multirow{4}{0.2 cm}{$Q_\mathrm{d}$} & -1 & 0 & 0 & 380 & 19 & \multirow{4}{0.2 cm}{$Q_\mathrm{d}$} & -1 & 0 & 0 & 380 & 19 \\ 
    {} & 0 & 0 & 0 & 0 & 1 & {} & 0 & 0 & 0 & 0 & 20 & {} & 0 & 0 & 0 & 0 & 10 \\ 
    {} & +1 & 0 & 0 & 0 & 0 & {} & +1 & 0 & 0 & 0 & 0 & {} & +1 & 0 & 0 & 0 & 0 \\ 
    {} & +2 & 0 & 0 & 0 & 0 & {} & +2 & 0 & 0 & 0 & 0 & {} & +2 & 0 & 0 & 0 & 0 \\ 
    \multicolumn{18}{c}{} \vspace{0.05 cm}\\
    \hline

    \multicolumn{18}{c}{Range-separated} \vspace{0.1 cm}\\
    \multicolumn{2}{c}{CAM-B3LYP} & \multicolumn{4}{c}{$Q_\mathrm{a}$} & \multicolumn{2}{c}{$\omega$B97} & \multicolumn{4}{c}{$Q_\mathrm{a}$} & \multicolumn{2}{c}{$\omega$PBE} & \multicolumn{4}{c}{$Q_\mathrm{a}$} \\ 
    \multicolumn{2}{c}{} & -1 & 0 & +1 & +2 & \multicolumn{2}{c}{} & -1 & 0 & +1 & +2 & \multicolumn{2}{c}{} & -1 & 0 & +1 & +2 \\ \cmidrule(lr){3-6} \cmidrule(lr){9-12} \cmidrule(lr){15-18}
    \multirow{4}{0.2 cm}{$Q_\mathrm{d}$} & -1 & 0 & 0 & 374 & 19 & \multirow{4}{0.2 cm}{$Q_\mathrm{d}$} & -1 & 0 & 0 & 0 & 17 & \multirow{4}{0.2 cm}{$Q_\mathrm{d}$} & -1 & 0 & 0 & 0 & 19 \\ 
    {} & 0 & 0 & 0 & 0 & 0 & {} & 0 & 0 & 0 & 0 & 0 & {} & 0 & 0 & 0 & 0 & 0 \\ 
    {} & +1 & 0 & 0 & 0 & 0 & {} & +1 & 0 & 0 & 0 & 0 & {} & +1 & 0 & 0 & 0 & 0 \\ 
    {} & +2 & 0 & 0 & 0 & 0 & {} & +2 & 0 & 0 & 0 & 0 & {} & +2 & 0 & 0 & 0 & 0 \\ 
    \multicolumn{18}{c}{} \\
    
    \bottomrule
    \bottomrule
    \end{tabular}
    \begin{tablenotes}
    \item $^a$ MN15 is both a hybrid functional and a meta functional, but we list it under ``Hybrid".
    \end{tablenotes}
    \caption{Number of non-Aufbau states predicted by various KS functionals for noninteracting pairs of the 20 natural amino acids in their neutral/protonated/deprotonated states. The data is broken down by charges $Q_\mathrm{d}$/$Q_\mathrm{a}$ of the ``donor''/``acceptor'' fragments, respectively (i.e., containing the HOMO and LUMO, respectively).}
    \label{table:non-aufbau_systems}
\end{table}

As expected, including larger fractions of Hartree-Fock exchange reduces the number of problematic combinations.
The performance of the semilocal functionals is almost unchanged going from LDA to GGA.
The meta-GGAs provide some improvement, but not as much improvement as hybrid functionals.
Finally, the range-separated functionals perform best of all, however, there are some cases where they still predict non-Aufbau filled systems.
The most difficult systems are when the donor is negatively charged and the acceptor has a +2 charge (Arginine), where even $\omega$B97 fails.

\begin{figure}

    \centering
    \includegraphics[width=6.25in]{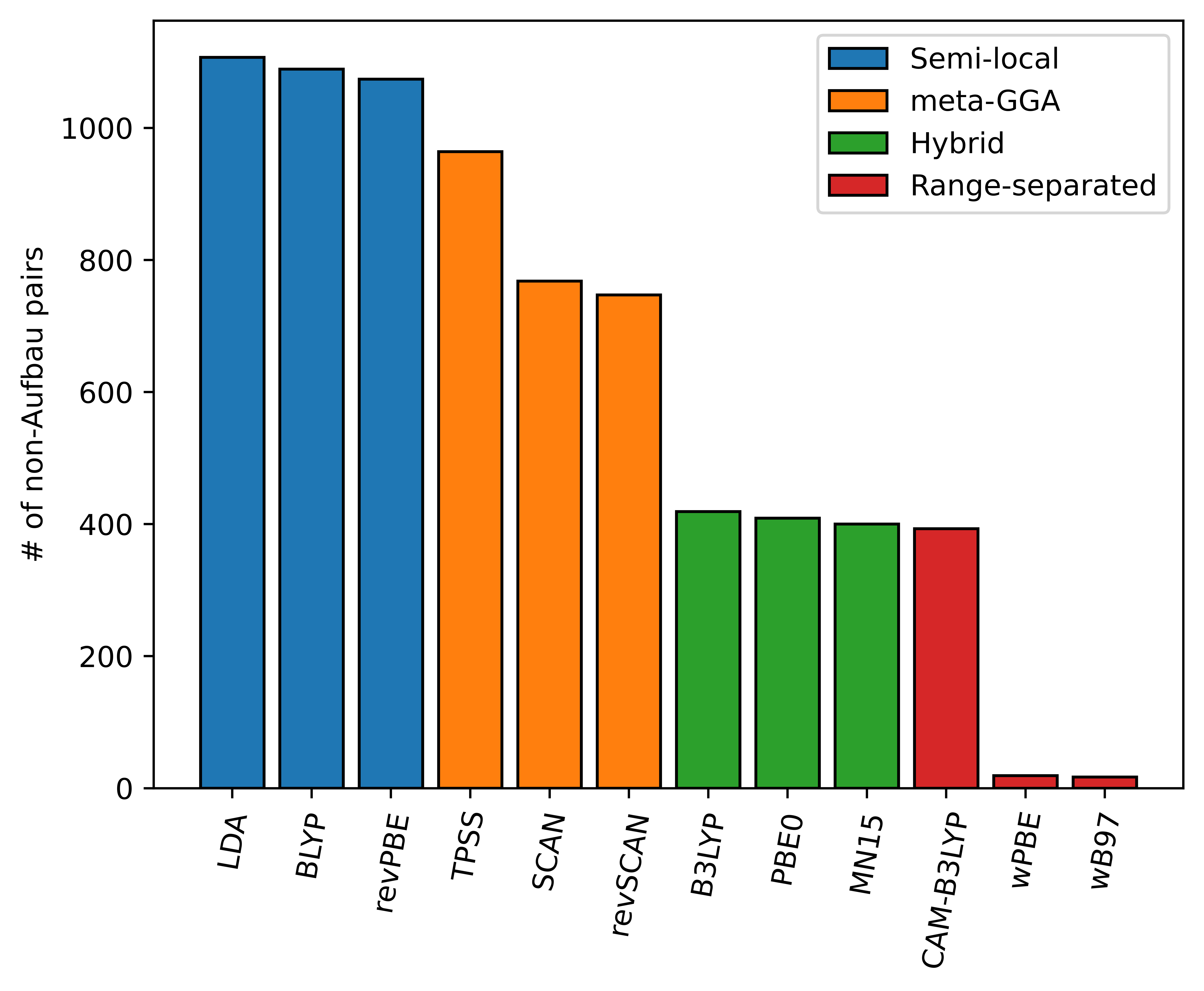}
    \caption{Overall number of non-Aufbau states predicted by various KS functionals for noninteracting pairs of the 20 natural amino acids in their neutral/protonated/deprotonated states. See \cref{table:non-aufbau_systems} for the further breakdown by the aminoacid charges.}
    \label{fig:nonaufbaupairs}
\end{figure}

\section{Summary}
\label{section:summary}

This work reexamined the SCF convergence problems for polypeptides in gas phase in conjunction with modern non-hybrid and hybrid DFAs.
While standard SCF solvers typically fail catastrophically,\cite{VRG:rudberg:2012:JPCM,VRG:ren:2022:JCP} using a robust quasi-Newton SCF solver\cite{VRG:slattery:2024:PCCP} we were able to obtain SCF solutions when the conventional solvers fail; in such cases the KS solutions always had a vanishing HOMO-LUMO gap.
Deeper analysis of these solutions using novel natural deformation orbitals obtained from the HF-KS density matrix difference reveals which regions of the system donate and accept electron density in the unphysical KS DFT solution, and are thus the culprits in the delocalization error.
In general the unphysical charge delocalization can involve not only charged moieties but also formally neutral fragments (e.g. a phenyl ring), demonstrating that zwitterions are not the only problematic cases for semilocal DFAs. The origin of the unphysical charge delocalization is the misalignment of the KS Fock operator eigenspectrum between molecular fragments. A systematic study of pairs of 20 naturally occurring amino acids in various protonation states suggested that the unphysical charge delocalization is only partially reduced by the use of a more sophisticated functional.
The range-separated functionals employing 100\% exact exchange at long range were a nearly perfect remedy, albeit not totally immune from the problem. The rest of the representative functional families (standard hybrid, meta-GGA, GGA) all suffered from the unphysical charge delocalization to various extents.

The most direct lesson from our work is the need for caution when unexpectedly small (or vanishing) HOMO-LUMO gaps and atypical SCF convergence patterns (e.g., oscillatory) are observed in KS DFT simulations in any context (bio or otherwise). Such anomalies should call for probing the solution (e.g., population analysis) and trying more advanced KS functionals.
Although our work focused on a specific, and somewhat artificial, biosimulation context, namely isolated polypeptides, such systems continue to serve as components and even as the sole focus of benchmark datasets\cite{VRG:prasad:2019:SD} for training more approximate models.
There are also lessons here for the broader class of Kohn-Sham DFT simulations. As affordable source of first-principles potentials for training more approximate models, KS DFT is increasingly used to generate massive datasets for benchmarking and training purposes,\cite{doi:10.1021/acs.jpcb.0c09251} with scale too large for validatation of even a nonnegligible portion of the dataset. So as the role of KS DFT as the ``ground truth'' model rises so should our expectations of its accuracy and robustness. Although we can expect that the continuing improvement of KS functionals will reduce the occurrence of the artifacts discussed here, the need for robust solvers will only rise as the degree of automation of KS DFT-based workflows continues to rise. The 

\section*{Acknowledgements}
This work was supported by the U.S. Department of Energy via award DE-SC0022327. We also acknowledge Advanced Research Computing at Virginia Tech (www.arc.vt.edu) for providing computational resources and technical support that have contributed to the results reported within this paper.

\bibliography{sas-refs,vrgrefs,additional_refs}

\providecommand{\latin}[1]{#1}
\makeatletter
\providecommand{\doi}
  {\begingroup\let\do\@makeother\dospecials
  \catcode`\{=1 \catcode`\}=2 \doi@aux}
\providecommand{\doi@aux}[1]{\endgroup\texttt{#1}}
\makeatother
\providecommand*\mcitethebibliography{\thebibliography}
\csname @ifundefined\endcsname{endmcitethebibliography}
  {\let\endmcitethebibliography\endthebibliography}{}
\begin{mcitethebibliography}{82}
\providecommand*\natexlab[1]{#1}
\providecommand*\mciteSetBstSublistMode[1]{}
\providecommand*\mciteSetBstMaxWidthForm[2]{}
\providecommand*\mciteBstWouldAddEndPuncttrue
  {\def\EndOfBibitem{\unskip.}}
\providecommand*\mciteBstWouldAddEndPunctfalse
  {\let\EndOfBibitem\relax}
\providecommand*\mciteSetBstMidEndSepPunct[3]{}
\providecommand*\mciteSetBstSublistLabelBeginEnd[3]{}
\providecommand*\EndOfBibitem{}
\mciteSetBstSublistMode{f}
\mciteSetBstMaxWidthForm{subitem}{(\alph{mcitesubitemcount})}
\mciteSetBstSublistLabelBeginEnd
  {\mcitemaxwidthsubitemform\space}
  {\relax}
  {\relax}

\bibitem[Tsuneda \latin{et~al.}(2010)Tsuneda, Song, Suzuki, and
  Hirao]{VRG:tsuneda:2010:JCP}
Tsuneda,~T.; Song,~J.-W.; Suzuki,~S.; Hirao,~K. On {{Koopmans}}' Theorem in
  Density Functional Theory. \emph{The Journal of Chemical Physics}
  \textbf{2010}, \emph{133}, 174101\relax
\mciteBstWouldAddEndPuncttrue
\mciteSetBstMidEndSepPunct{\mcitedefaultmidpunct}
{\mcitedefaultendpunct}{\mcitedefaultseppunct}\relax
\EndOfBibitem
\bibitem[Rudberg \latin{et~al.}(2011)Rudberg, Rubensson, and
  Sa{\l}ek]{VRG:rudberg:2011:JCTC}
Rudberg,~E.; Rubensson,~E.~H.; Sa{\l}ek,~P. {{Kohn}}-{{Sham Density Functional
  Theory Electronic Structure Calculations}} with {{Linearly Scaling
  Computational Time}} and {{Memory Usage}}. \emph{J. Chem. Theory Comput.}
  \textbf{2011}, \emph{7}, 340--350\relax
\mciteBstWouldAddEndPuncttrue
\mciteSetBstMidEndSepPunct{\mcitedefaultmidpunct}
{\mcitedefaultendpunct}{\mcitedefaultseppunct}\relax
\EndOfBibitem
\bibitem[Tozer(2003)]{VRG:tozer:2003:JCP}
Tozer,~D.~J. Relationship between Long-Range Charge-Transfer Excitation Energy
  Error and Integer Discontinuity in {{Kohn}}--{{Sham}} Theory. \emph{J. Chem.
  Phys.} \textbf{2003}, \emph{119}, 12697--12699\relax
\mciteBstWouldAddEndPuncttrue
\mciteSetBstMidEndSepPunct{\mcitedefaultmidpunct}
{\mcitedefaultendpunct}{\mcitedefaultseppunct}\relax
\EndOfBibitem
\bibitem[Dreuw \latin{et~al.}(2003)Dreuw, Weisman, and
  {Head-Gordon}]{VRG:dreuw:2003:JCP}
Dreuw,~A.; Weisman,~J.~L.; {Head-Gordon},~M. Long-Range Charge-Transfer Excited
  States in Time-Dependent Density Functional Theory Require Non-Local
  Exchange. \emph{J. Chem. Phys.} \textbf{2003}, \emph{119}, 2943--2946\relax
\mciteBstWouldAddEndPuncttrue
\mciteSetBstMidEndSepPunct{\mcitedefaultmidpunct}
{\mcitedefaultendpunct}{\mcitedefaultseppunct}\relax
\EndOfBibitem
\bibitem[Casida \latin{et~al.}(1998)Casida, Jamorski, Casida, and
  Salahub]{VRG:casida:1998:JCP}
Casida,~M.~E.; Jamorski,~C.; Casida,~K.~C.; Salahub,~D.~R. Molecular Excitation
  Energies to High-Lying Bound States from Time-Dependent Density-Functional
  Response Theory: {{Characterization}} and Correction of the Time-Dependent
  Local Density Approximation Ionization Threshold. \emph{J. Chem. Phys.}
  \textbf{1998}, \emph{108}, 4439--4449\relax
\mciteBstWouldAddEndPuncttrue
\mciteSetBstMidEndSepPunct{\mcitedefaultmidpunct}
{\mcitedefaultendpunct}{\mcitedefaultseppunct}\relax
\EndOfBibitem
\bibitem[Baerends \latin{et~al.}(2013)Baerends, Gritsenko, and
  Van~Meer]{VRG:baerends:2013:PCCP}
Baerends,~E.~J.; Gritsenko,~O.~V.; Van~Meer,~R. The {{Kohn}}--{{Sham}} Gap, the
  Fundamental Gap and the Optical Gap: The Physical Meaning of Occupied and
  Virtual {{Kohn}}--{{Sham}} Orbital Energies. \emph{Phys. Chem. Chem. Phys.}
  \textbf{2013}, \emph{15}, 16408\relax
\mciteBstWouldAddEndPuncttrue
\mciteSetBstMidEndSepPunct{\mcitedefaultmidpunct}
{\mcitedefaultendpunct}{\mcitedefaultseppunct}\relax
\EndOfBibitem
\bibitem[Jensen(2010)]{VRG:jensen:2010:JCTC}
Jensen,~F. Describing {{Anions}} by {{Density Functional Theory}}: {{Fractional
  Electron Affinity}}. \emph{J. Chem. Theory Comput.} \textbf{2010}, \emph{6},
  2726--2735\relax
\mciteBstWouldAddEndPuncttrue
\mciteSetBstMidEndSepPunct{\mcitedefaultmidpunct}
{\mcitedefaultendpunct}{\mcitedefaultseppunct}\relax
\EndOfBibitem
\bibitem[Peach \latin{et~al.}(2015)Peach, Teale, Helgaker, and
  Tozer]{VRG:peach:2015:JCTC}
Peach,~M. J.~G.; Teale,~A.~M.; Helgaker,~T.; Tozer,~D.~J. Fractional {{Electron
  Loss}} in {{Approximate DFT}} and {{Hartree}}--{{Fock Theory}}. \emph{J.
  Chem. Theory Comput.} \textbf{2015}, \emph{11}, 5262--5268\relax
\mciteBstWouldAddEndPuncttrue
\mciteSetBstMidEndSepPunct{\mcitedefaultmidpunct}
{\mcitedefaultendpunct}{\mcitedefaultseppunct}\relax
\EndOfBibitem
\bibitem[Dobbs and Dixon(1994)Dobbs, and Dixon]{VRG:dobbs:1994:JPC}
Dobbs,~K.~D.; Dixon,~D.~A. Ab {{Initio Prediction}} of the {{Activation
  Energy}} for the {{Abstraction}} of a {{Hydrogen Atom}} from {{Methane}} by
  {{Chlorine Atom}}. \emph{J. Phys. Chem.} \textbf{1994}, \emph{98},
  12584--12589\relax
\mciteBstWouldAddEndPuncttrue
\mciteSetBstMidEndSepPunct{\mcitedefaultmidpunct}
{\mcitedefaultendpunct}{\mcitedefaultseppunct}\relax
\EndOfBibitem
\bibitem[Johnson \latin{et~al.}(1994)Johnson, Gonzales, Gill, and
  Pople]{VRG:johnson:1994:CPL}
Johnson,~B.~G.; Gonzales,~C.~A.; Gill,~P.~M.; Pople,~J.~A. A Density Functional
  Study of the Simplest Hydrogen Abstraction Reaction. {{Effect}} of
  Self-Interaction Correction. \emph{Chemical Physics Letters} \textbf{1994},
  \emph{221}, 100--108\relax
\mciteBstWouldAddEndPuncttrue
\mciteSetBstMidEndSepPunct{\mcitedefaultmidpunct}
{\mcitedefaultendpunct}{\mcitedefaultseppunct}\relax
\EndOfBibitem
\bibitem[Zhang \latin{et~al.}(1995)Zhang, Bell, and Truong]{VRG:zhang:1995:JPC}
Zhang,~Q.; Bell,~R.; Truong,~T.~N. Ab {{Initio}} and {{Density Functional
  Theory Studies}} of {{Proton Transfer Reactions}} in {{Multiple Hydrogen Bond
  Systems}}. \emph{J. Phys. Chem.} \textbf{1995}, \emph{99}, 592--599\relax
\mciteBstWouldAddEndPuncttrue
\mciteSetBstMidEndSepPunct{\mcitedefaultmidpunct}
{\mcitedefaultendpunct}{\mcitedefaultseppunct}\relax
\EndOfBibitem
\bibitem[Vydrov and Scuseria(2006)Vydrov, and Scuseria]{VRG:vydrov:2006:JCPa}
Vydrov,~O.~A.; Scuseria,~G.~E. A Simple Method to Selectively Scale down the
  Self-Interaction Correction. \emph{J. Chem. Phys.} \textbf{2006}, \emph{124},
  191101\relax
\mciteBstWouldAddEndPuncttrue
\mciteSetBstMidEndSepPunct{\mcitedefaultmidpunct}
{\mcitedefaultendpunct}{\mcitedefaultseppunct}\relax
\EndOfBibitem
\bibitem[Janesko and Scuseria(2008)Janesko, and Scuseria]{VRG:janesko:2008:JCP}
Janesko,~B.~G.; Scuseria,~G.~E. Hartree--{{Fock}} Orbitals Significantly
  Improve the Reaction Barrier Heights Predicted by Semilocal Density
  Functionals. \emph{J. Chem. Phys.} \textbf{2008}, \emph{128}, 244112\relax
\mciteBstWouldAddEndPuncttrue
\mciteSetBstMidEndSepPunct{\mcitedefaultmidpunct}
{\mcitedefaultendpunct}{\mcitedefaultseppunct}\relax
\EndOfBibitem
\bibitem[Zhang and Yang(1998)Zhang, and Yang]{VRG:zhang:1998:JCP}
Zhang,~Y.; Yang,~W. A Challenge for Density Functionals: {{Self-interaction}}
  Error Increases for Systems with a Noninteger Number of Electrons. \emph{J.
  Chem. Phys.} \textbf{1998}, \emph{109}, 2604--2608\relax
\mciteBstWouldAddEndPuncttrue
\mciteSetBstMidEndSepPunct{\mcitedefaultmidpunct}
{\mcitedefaultendpunct}{\mcitedefaultseppunct}\relax
\EndOfBibitem
\bibitem[Dutoi and {Head-Gordon}(2006)Dutoi, and
  {Head-Gordon}]{VRG:dutoi:2006:CPL}
Dutoi,~A.~D.; {Head-Gordon},~M. Self-Interaction Error of Local Density
  Functionals for Alkali--Halide Dissociation. \emph{Chem. Phys. Lett.}
  \textbf{2006}, \emph{422}, 230--233\relax
\mciteBstWouldAddEndPuncttrue
\mciteSetBstMidEndSepPunct{\mcitedefaultmidpunct}
{\mcitedefaultendpunct}{\mcitedefaultseppunct}\relax
\EndOfBibitem
\bibitem[{Otero-de-la-Roza} and Johnson(2020){Otero-de-la-Roza}, and
  Johnson]{VRG:otero-de-la-roza:2020:JPCA}
{Otero-de-la-Roza},~A.; Johnson,~E.~R. Analysis of {{Density-Functional
  Errors}} for {{Noncovalent Interactions}} between {{Charged Molecules}}.
  \emph{J. Phys. Chem. A} \textbf{2020}, \emph{124}, 353--361\relax
\mciteBstWouldAddEndPuncttrue
\mciteSetBstMidEndSepPunct{\mcitedefaultmidpunct}
{\mcitedefaultendpunct}{\mcitedefaultseppunct}\relax
\EndOfBibitem
\bibitem[Ruiz \latin{et~al.}(1996)Ruiz, Salahub, and Vela]{VRG:ruiz:1996:JPC}
Ruiz,~E.; Salahub,~D.~R.; Vela,~A. Charge-{{Transfer Complexes}}: {{Stringent
  Tests}} for {{Widely Used Density Functionals}}. \emph{J. Phys. Chem.}
  \textbf{1996}, \emph{100}, 12265--12276\relax
\mciteBstWouldAddEndPuncttrue
\mciteSetBstMidEndSepPunct{\mcitedefaultmidpunct}
{\mcitedefaultendpunct}{\mcitedefaultseppunct}\relax
\EndOfBibitem
\bibitem[Isborn \latin{et~al.}(2013)Isborn, Mar, Curchod, Tavernelli, and
  Mart{\'i}nez]{VRG:isborn:2013:JPCB}
Isborn,~C.~M.; Mar,~B.~D.; Curchod,~B. F.~E.; Tavernelli,~I.;
  Mart{\'i}nez,~T.~J. The {{Charge Transfer Problem}} in {{Density Functional
  Theory Calculations}} of {{Aqueously Solvated Molecules}}. \emph{J. Phys.
  Chem. B} \textbf{2013}, \emph{117}, 12189--12201\relax
\mciteBstWouldAddEndPuncttrue
\mciteSetBstMidEndSepPunct{\mcitedefaultmidpunct}
{\mcitedefaultendpunct}{\mcitedefaultseppunct}\relax
\EndOfBibitem
\bibitem[{Otero-de-la-Roza} \latin{et~al.}(2014){Otero-de-la-Roza}, Johnson,
  and DiLabio]{VRG:otero-de-la-roza:2014:JCTC}
{Otero-de-la-Roza},~A.; Johnson,~E.~R.; DiLabio,~G.~A. Halogen {{Bonding}} from
  {{Dispersion-Corrected Density-Functional Theory}}: {{The Role}} of
  {{Delocalization Error}}. \emph{J. Chem. Theory Comput.} \textbf{2014},
  \emph{10}, 5436--5447\relax
\mciteBstWouldAddEndPuncttrue
\mciteSetBstMidEndSepPunct{\mcitedefaultmidpunct}
{\mcitedefaultendpunct}{\mcitedefaultseppunct}\relax
\EndOfBibitem
\bibitem[Rubensson and Rudberg(2011)Rubensson, and
  Rudberg]{VRG:rubensson:2011:JCC}
Rubensson,~E.~H.; Rudberg,~E. Bringing about Matrix Sparsity in Linear-scaling
  Electronic Structure Calculations. \emph{J Comput Chem} \textbf{2011},
  \emph{32}, 1411--1423\relax
\mciteBstWouldAddEndPuncttrue
\mciteSetBstMidEndSepPunct{\mcitedefaultmidpunct}
{\mcitedefaultendpunct}{\mcitedefaultseppunct}\relax
\EndOfBibitem
\bibitem[Rudberg(2012)]{VRG:rudberg:2012:JPCM}
Rudberg,~E. Difficulties in Applying Pure {{Kohn}}--{{Sham}} Density Functional
  Theory Electronic Structure Methods to Protein Molecules. \emph{J. Phys.:
  Condens. Matter} \textbf{2012}, \emph{24}, 072202\relax
\mciteBstWouldAddEndPuncttrue
\mciteSetBstMidEndSepPunct{\mcitedefaultmidpunct}
{\mcitedefaultendpunct}{\mcitedefaultseppunct}\relax
\EndOfBibitem
\bibitem[Antony and Grimme(2012)Antony, and Grimme]{VRG:antony:2012:JCC}
Antony,~J.; Grimme,~S. Fully {\emph{Ab Initio}} Protein-ligand Interaction
  Energies with Dispersion Corrected Density Functional Theory. \emph{J Comput
  Chem} \textbf{2012}, \emph{33}, 1730--1739\relax
\mciteBstWouldAddEndPuncttrue
\mciteSetBstMidEndSepPunct{\mcitedefaultmidpunct}
{\mcitedefaultendpunct}{\mcitedefaultseppunct}\relax
\EndOfBibitem
\bibitem[Kulik \latin{et~al.}(2012)Kulik, Luehr, Ufimtsev, and
  Martinez]{VRG:kulik:2012:JPCB}
Kulik,~H.~J.; Luehr,~N.; Ufimtsev,~I.~S.; Martinez,~T.~J. Ab {{Initio Quantum
  Chemistry}} for {{Protein Structures}}. \emph{J. Phys. Chem. B}
  \textbf{2012}, \emph{116}, 12501--12509\relax
\mciteBstWouldAddEndPuncttrue
\mciteSetBstMidEndSepPunct{\mcitedefaultmidpunct}
{\mcitedefaultendpunct}{\mcitedefaultseppunct}\relax
\EndOfBibitem
\bibitem[Lever \latin{et~al.}(2013)Lever, Cole, Hine, Haynes, and
  Payne]{VRG:lever:2013:JPCM}
Lever,~G.; Cole,~D.~J.; Hine,~N. D.~M.; Haynes,~P.~D.; Payne,~M.~C.
  Electrostatic Considerations Affecting the Calculated {{HOMO}}--{{LUMO}} Gap
  in Protein Molecules. \emph{J. Phys.: Condens. Matter} \textbf{2013},
  \emph{25}, 152101\relax
\mciteBstWouldAddEndPuncttrue
\mciteSetBstMidEndSepPunct{\mcitedefaultmidpunct}
{\mcitedefaultendpunct}{\mcitedefaultseppunct}\relax
\EndOfBibitem
\bibitem[Cole and Hine(2016)Cole, and Hine]{VRG:cole:2016:JPCM}
Cole,~D.~J.; Hine,~N. D.~M. Applications of Large-Scale Density Functional
  Theory in Biology. \emph{J. Phys.: Condens. Matter} \textbf{2016}, \emph{28},
  393001\relax
\mciteBstWouldAddEndPuncttrue
\mciteSetBstMidEndSepPunct{\mcitedefaultmidpunct}
{\mcitedefaultendpunct}{\mcitedefaultseppunct}\relax
\EndOfBibitem
\bibitem[Zuehlsdorff \latin{et~al.}(2016)Zuehlsdorff, Haynes, Hanke, Payne, and
  Hine]{VRG:zuehlsdorff:2016:JCTC}
Zuehlsdorff,~T.~J.; Haynes,~P.~D.; Hanke,~F.; Payne,~M.~C.; Hine,~N. D.~M.
  Solvent {{Effects}} on {{Electronic Excitations}} of an {{Organic
  Chromophore}}. \emph{J. Chem. Theory Comput.} \textbf{2016}, \emph{12},
  1853--1861\relax
\mciteBstWouldAddEndPuncttrue
\mciteSetBstMidEndSepPunct{\mcitedefaultmidpunct}
{\mcitedefaultendpunct}{\mcitedefaultseppunct}\relax
\EndOfBibitem
\bibitem[Ren and Liu(2022)Ren, and Liu]{VRG:ren:2022:JCP}
Ren,~F.; Liu,~F. Impacts of Polarizable Continuum Models on the {{SCF}}
  Convergence and {{DFT}} Delocalization Error of Large Molecules. \emph{J.
  Chem. Phys.} \textbf{2022}, \emph{157}, 184106\relax
\mciteBstWouldAddEndPuncttrue
\mciteSetBstMidEndSepPunct{\mcitedefaultmidpunct}
{\mcitedefaultendpunct}{\mcitedefaultseppunct}\relax
\EndOfBibitem
\bibitem[Sepunaru \latin{et~al.}(2015)Sepunaru, {Refaely-Abramson}, Lovrin{\v
  c}i{\'c}, Gavrilov, Agrawal, Levy, Kronik, Pecht, Sheves, and
  Cahen]{VRG:sepunaru:2015:JACS}
Sepunaru,~L.; {Refaely-Abramson},~S.; Lovrin{\v c}i{\'c},~R.; Gavrilov,~Y.;
  Agrawal,~P.; Levy,~Y.; Kronik,~L.; Pecht,~I.; Sheves,~M.; Cahen,~D.
  Electronic {{Transport}} via {{Homopeptides}}: {{The Role}} of {{Side
  Chains}} and {{Secondary Structure}}. \emph{J. Am. Chem. Soc.} \textbf{2015},
  \emph{137}, 9617--9626\relax
\mciteBstWouldAddEndPuncttrue
\mciteSetBstMidEndSepPunct{\mcitedefaultmidpunct}
{\mcitedefaultendpunct}{\mcitedefaultseppunct}\relax
\EndOfBibitem
\bibitem[Sharley(2016)]{VRG:sharley:2016:}
Sharley,~J.~N. Amino Acid Preference against Beta Sheet through Allowing
  Backbone Hydration Enabled by the Presence of Cation. 2016\relax
\mciteBstWouldAddEndPuncttrue
\mciteSetBstMidEndSepPunct{\mcitedefaultmidpunct}
{\mcitedefaultendpunct}{\mcitedefaultseppunct}\relax
\EndOfBibitem
\bibitem[Vydrov and Scuseria(2006)Vydrov, and Scuseria]{VRG:vydrov:2006:JCPb}
Vydrov,~O.~A.; Scuseria,~G.~E. Assessment of a Long-Range Corrected Hybrid
  Functional. \emph{J. Chem. Phys.} \textbf{2006}, \emph{125}, 234109\relax
\mciteBstWouldAddEndPuncttrue
\mciteSetBstMidEndSepPunct{\mcitedefaultmidpunct}
{\mcitedefaultendpunct}{\mcitedefaultseppunct}\relax
\EndOfBibitem
\bibitem[Li \latin{et~al.}(2015)Li, Zuehlsdorff, Payne, and
  Hine]{VRG:li:2015:PCCP}
Li,~J.-H.; Zuehlsdorff,~T.~J.; Payne,~M.~C.; Hine,~N. D.~M. Identifying and
  Tracing Potential Energy Surfaces of Electronic Excitations with Specific
  Character via Their Transition Origins: Application to Oxirane. \emph{Phys.
  Chem. Chem. Phys.} \textbf{2015}, \emph{17}, 12065--12079\relax
\mciteBstWouldAddEndPuncttrue
\mciteSetBstMidEndSepPunct{\mcitedefaultmidpunct}
{\mcitedefaultendpunct}{\mcitedefaultseppunct}\relax
\EndOfBibitem
\bibitem[Shore \latin{et~al.}(1977)Shore, Rose, and
  Zaremba]{VRG:shore:1977:PRB}
Shore,~H.~B.; Rose,~J.~H.; Zaremba,~E. Failure of the Local Exchange
  Approximation in the Evaluation of the {{H}} - Ground State. \emph{Phys. Rev.
  B} \textbf{1977}, \emph{15}, 2858--2861\relax
\mciteBstWouldAddEndPuncttrue
\mciteSetBstMidEndSepPunct{\mcitedefaultmidpunct}
{\mcitedefaultendpunct}{\mcitedefaultseppunct}\relax
\EndOfBibitem
\bibitem[{K Schwarz}(1978)]{VRG:kschwarz:1978:JPBAMP}
{K Schwarz}, First Ionisation Potentials of Atoms Obtained with Local-Density
  Schemes. \emph{J. Phys. B: Atom. Mol. Phys.} \textbf{1978}, \emph{11},
  1339--1351\relax
\mciteBstWouldAddEndPuncttrue
\mciteSetBstMidEndSepPunct{\mcitedefaultmidpunct}
{\mcitedefaultendpunct}{\mcitedefaultseppunct}\relax
\EndOfBibitem
\bibitem[Schwarz(1978)]{VRG:schwarz:1978:CPL}
Schwarz,~K. Instability of Stable Negative Ions in the {{X$\alpha$}} Method or
  Other Local Density Functional Schemes. \emph{Chemical Physics Letters}
  \textbf{1978}, \emph{57}, 605--607\relax
\mciteBstWouldAddEndPuncttrue
\mciteSetBstMidEndSepPunct{\mcitedefaultmidpunct}
{\mcitedefaultendpunct}{\mcitedefaultseppunct}\relax
\EndOfBibitem
\bibitem[R{\"o}sch and Trickey(1997)R{\"o}sch, and Trickey]{VRG:rosch:1997:JCP}
R{\"o}sch,~N.; Trickey,~S.~B. Comment on ``{{Concerning}} the Applicability of
  Density Functional Methods to Atomic and Molecular Negative Ions'' [{{J}}.
  {{Chem}}. {{Phys}}. {\textbf{105}} , 862 (1996)]. \emph{J. Chem. Phys.}
  \textbf{1997}, \emph{106}, 8940--8941\relax
\mciteBstWouldAddEndPuncttrue
\mciteSetBstMidEndSepPunct{\mcitedefaultmidpunct}
{\mcitedefaultendpunct}{\mcitedefaultseppunct}\relax
\EndOfBibitem
\bibitem[Peach \latin{et~al.}(2010)Peach, De~Proft, and
  Tozer]{VRG:peach:2010:JPCL}
Peach,~M. J.~G.; De~Proft,~F.; Tozer,~D.~J. Negative {{Electron Affinities}}
  from {{DFT}}: {{Fluorination}} of {{Ethylene}}. \emph{J. Phys. Chem. Lett.}
  \textbf{2010}, \emph{1}, 2826--2831\relax
\mciteBstWouldAddEndPuncttrue
\mciteSetBstMidEndSepPunct{\mcitedefaultmidpunct}
{\mcitedefaultendpunct}{\mcitedefaultseppunct}\relax
\EndOfBibitem
\bibitem[Lee \latin{et~al.}(2010)Lee, Furche, and Burke]{VRG:lee:2010:JPCL}
Lee,~D.; Furche,~F.; Burke,~K. Accuracy of {{Electron Affinities}} of {{Atoms}}
  in {{Approximate Density Functional Theory}}. \emph{J. Phys. Chem. Lett.}
  \textbf{2010}, \emph{1}, 2124--2129\relax
\mciteBstWouldAddEndPuncttrue
\mciteSetBstMidEndSepPunct{\mcitedefaultmidpunct}
{\mcitedefaultendpunct}{\mcitedefaultseppunct}\relax
\EndOfBibitem
\bibitem[Kim \latin{et~al.}(2011)Kim, Sim, and Burke]{VRG:kim:2011:JCP}
Kim,~M.-C.; Sim,~E.; Burke,~K. Communication: {{Avoiding}} Unbound Anions in
  Density Functional Calculations. \emph{J. Chem. Phys.} \textbf{2011},
  \emph{134}, 171103\relax
\mciteBstWouldAddEndPuncttrue
\mciteSetBstMidEndSepPunct{\mcitedefaultmidpunct}
{\mcitedefaultendpunct}{\mcitedefaultseppunct}\relax
\EndOfBibitem
\bibitem[{Mori-S{\'a}nchez} \latin{et~al.}(2006){Mori-S{\'a}nchez}, Cohen, and
  Yang]{VRG:mori-sanchez:2006:JCP}
{Mori-S{\'a}nchez},~P.; Cohen,~A.~J.; Yang,~W. Many-Electron Self-Interaction
  Error in Approximate Density Functionals. \emph{J. Chem. Phys.}
  \textbf{2006}, \emph{125}, 201102\relax
\mciteBstWouldAddEndPuncttrue
\mciteSetBstMidEndSepPunct{\mcitedefaultmidpunct}
{\mcitedefaultendpunct}{\mcitedefaultseppunct}\relax
\EndOfBibitem
\bibitem[{Mori-S{\'a}nchez} \latin{et~al.}(2008){Mori-S{\'a}nchez}, Cohen, and
  Yang]{VRG:mori-sanchez:2008:PRL}
{Mori-S{\'a}nchez},~P.; Cohen,~A.~J.; Yang,~W. Localization and
  {{Delocalization Errors}} in {{Density Functional Theory}} and
  {{Implications}} for {{Band-Gap Prediction}}. \emph{Phys. Rev. Lett.}
  \textbf{2008}, \emph{100}, 146401\relax
\mciteBstWouldAddEndPuncttrue
\mciteSetBstMidEndSepPunct{\mcitedefaultmidpunct}
{\mcitedefaultendpunct}{\mcitedefaultseppunct}\relax
\EndOfBibitem
\bibitem[Cohen \latin{et~al.}(2008)Cohen, {Mori-S{\'a}nchez}, and
  Yang]{VRG:cohen:2008:S}
Cohen,~A.~J.; {Mori-S{\'a}nchez},~P.; Yang,~W. Insights into {{Current
  Limitations}} of {{Density Functional Theory}}. \emph{Science} \textbf{2008},
  \emph{321}, 792--794\relax
\mciteBstWouldAddEndPuncttrue
\mciteSetBstMidEndSepPunct{\mcitedefaultmidpunct}
{\mcitedefaultendpunct}{\mcitedefaultseppunct}\relax
\EndOfBibitem
\bibitem[Perdew \latin{et~al.}(1982)Perdew, Parr, Levy, and
  Balduz]{VRG:perdew:1982:PRL}
Perdew,~J.~P.; Parr,~R.~G.; Levy,~M.; Balduz,~J.~L. Density-{{Functional
  Theory}} for {{Fractional Particle Number}}: {{Derivative Discontinuities}}
  of the {{Energy}}. \emph{Phys. Rev. Lett.} \textbf{1982}, \emph{49},
  1691--1694\relax
\mciteBstWouldAddEndPuncttrue
\mciteSetBstMidEndSepPunct{\mcitedefaultmidpunct}
{\mcitedefaultendpunct}{\mcitedefaultseppunct}\relax
\EndOfBibitem
\bibitem[Hait and {Head-Gordon}(2018)Hait, and
  {Head-Gordon}]{VRG:hait:2018:JPCL}
Hait,~D.; {Head-Gordon},~M. Delocalization {{Errors}} in {{Density Functional
  Theory Are Essentially Quadratic}} in {{Fractional Occupation Number}}.
  \emph{J. Phys. Chem. Lett.} \textbf{2018}, \emph{9}, 6280--6288\relax
\mciteBstWouldAddEndPuncttrue
\mciteSetBstMidEndSepPunct{\mcitedefaultmidpunct}
{\mcitedefaultendpunct}{\mcitedefaultseppunct}\relax
\EndOfBibitem
\bibitem[Jakobsen \latin{et~al.}(2013)Jakobsen, Kristensen, and
  Jensen]{VRG:jakobsen:2013:JCTC}
Jakobsen,~S.; Kristensen,~K.; Jensen,~F. Electrostatic {{Potential}} of
  {{Insulin}}: {{Exploring}} the {{Limitations}} of {{Density Functional
  Theory}} and {{Force Field Methods}}. \emph{J. Chem. Theory Comput.}
  \textbf{2013}, \emph{9}, 3978--3985\relax
\mciteBstWouldAddEndPuncttrue
\mciteSetBstMidEndSepPunct{\mcitedefaultmidpunct}
{\mcitedefaultendpunct}{\mcitedefaultseppunct}\relax
\EndOfBibitem
\bibitem[Sim \latin{et~al.}(2022)Sim, Song, Vuckovic, and
  Burke]{VRG:sim:2022:JACS}
Sim,~E.; Song,~S.; Vuckovic,~S.; Burke,~K. Improving {{Results}} by {{Improving
  Densities}}: {{Density-Corrected Density Functional Theory}}. \emph{J. Am.
  Chem. Soc.} \textbf{2022}, \emph{144}, 6625--6639\relax
\mciteBstWouldAddEndPuncttrue
\mciteSetBstMidEndSepPunct{\mcitedefaultmidpunct}
{\mcitedefaultendpunct}{\mcitedefaultseppunct}\relax
\EndOfBibitem
\bibitem[Slattery \latin{et~al.}(2024)Slattery, Surjuse, Peterson, Penchoff,
  and Valeev]{VRG:slattery:2024:PCCP}
Slattery,~S.~A.; Surjuse,~K.~A.; Peterson,~C.~C.; Penchoff,~D.~A.;
  Valeev,~E.~F. Economical Quasi-{{Newton}} Unitary Optimization of Electronic
  Orbitals. \emph{Phys. Chem. Chem. Phys.} \textbf{2024}, \emph{26},
  6557--6573\relax
\mciteBstWouldAddEndPuncttrue
\mciteSetBstMidEndSepPunct{\mcitedefaultmidpunct}
{\mcitedefaultendpunct}{\mcitedefaultseppunct}\relax
\EndOfBibitem
\bibitem[Pakiari \latin{et~al.}(2008)Pakiari, Fakhraee, and
  Azami]{VRG:pakiari:2008:IJQC}
Pakiari,~A.~H.; Fakhraee,~S.; Azami,~S.~M. Decomposition of Deformation Density
  into Orbital Components. \emph{Int. J. Quantum Chem.} \textbf{2008},
  \emph{108}, 415--422\relax
\mciteBstWouldAddEndPuncttrue
\mciteSetBstMidEndSepPunct{\mcitedefaultmidpunct}
{\mcitedefaultendpunct}{\mcitedefaultseppunct}\relax
\EndOfBibitem
\bibitem[Peng \latin{et~al.}(2020)Peng, Lewis, Wang, Clement, Pierce, Rishi,
  Pavo{\v s}evi{\'c}, Slattery, Zhang, Teke, Kumar, Masteran, Asadchev, Calvin,
  and Valeev]{VRG:peng:2020:JCP}
Peng,~C.; Lewis,~C.~A.; Wang,~X.; Clement,~M.~C.; Pierce,~K.; Rishi,~V.;
  Pavo{\v s}evi{\'c},~F.; Slattery,~S.; Zhang,~J.; Teke,~N.; Kumar,~A.;
  Masteran,~C.; Asadchev,~A.; Calvin,~J.~A.; Valeev,~E.~F. Massively {{Parallel
  Quantum Chemistry}}: {{A}} High-Performance Research Platform for Electronic
  Structure. \emph{J. Chem. Phys.} \textbf{2020}, \emph{153}, 044120\relax
\mciteBstWouldAddEndPuncttrue
\mciteSetBstMidEndSepPunct{\mcitedefaultmidpunct}
{\mcitedefaultendpunct}{\mcitedefaultseppunct}\relax
\EndOfBibitem
\bibitem[Petrone \latin{et~al.}(2018)Petrone, {Williams-Young}, Sun, Stetina,
  and Li]{VRG:petrone:2018:EPJB}
Petrone,~A.; {Williams-Young},~D.~B.; Sun,~S.; Stetina,~T.~F.; Li,~X. An
  Efficient Implementation of Two-Component Relativistic Density Functional
  Theory with Torque-Free Auxiliary Variables. \emph{Eur. Phys. J. B}
  \textbf{2018}, \emph{91}, 169\relax
\mciteBstWouldAddEndPuncttrue
\mciteSetBstMidEndSepPunct{\mcitedefaultmidpunct}
{\mcitedefaultendpunct}{\mcitedefaultseppunct}\relax
\EndOfBibitem
\bibitem[Lehtola \latin{et~al.}(2018)Lehtola, Steigemann, Oliveira, and
  Marques]{VRG:lehtola:2018:S}
Lehtola,~S.; Steigemann,~C.; Oliveira,~M.~J.; Marques,~M.~A. Recent
  Developments in Libxc --- {{A}} Comprehensive Library of Functionals for
  Density Functional Theory. \emph{SoftwareX} \textbf{2018}, \emph{7},
  1--5\relax
\mciteBstWouldAddEndPuncttrue
\mciteSetBstMidEndSepPunct{\mcitedefaultmidpunct}
{\mcitedefaultendpunct}{\mcitedefaultseppunct}\relax
\EndOfBibitem
\bibitem[Mura and Knowles(1996)Mura, and Knowles]{VRG:mura:1996:JCP}
Mura,~M.~E.; Knowles,~P.~J. Improved Radial Grids for Quadrature in Molecular
  Density-Functional Calculations. \emph{J. Chem. Phys.} \textbf{1996},
  \emph{104}, 9848--9858\relax
\mciteBstWouldAddEndPuncttrue
\mciteSetBstMidEndSepPunct{\mcitedefaultmidpunct}
{\mcitedefaultendpunct}{\mcitedefaultseppunct}\relax
\EndOfBibitem
\bibitem[Lebedev and Laikov(1999)Lebedev, and Laikov]{VRG:lebedev:1999:DM}
Lebedev,~V.~I.; Laikov,~D.~N. A Quadrature Formula for the Sphere of the 131st
  Algebraic Order of Accuracy. \emph{Dokl. Math.} \textbf{1999}, \emph{59},
  477--481\relax
\mciteBstWouldAddEndPuncttrue
\mciteSetBstMidEndSepPunct{\mcitedefaultmidpunct}
{\mcitedefaultendpunct}{\mcitedefaultseppunct}\relax
\EndOfBibitem
\bibitem[Dirac(1930)]{VRG:dirac:1930:MPCPS}
Dirac,~P. A.~M. Note on {{Exchange Phenomena}} in the {{Thomas Atom}}.
  \emph{Math. Proc. Camb. Phil. Soc.} \textbf{1930}, \emph{26}, 376--385\relax
\mciteBstWouldAddEndPuncttrue
\mciteSetBstMidEndSepPunct{\mcitedefaultmidpunct}
{\mcitedefaultendpunct}{\mcitedefaultseppunct}\relax
\EndOfBibitem
\bibitem[Vosko \latin{et~al.}(1980)Vosko, Wilk, and Nusair]{VRG:vosko:1980:CJP}
Vosko,~S.~H.; Wilk,~L.; Nusair,~M. Accurate Spin-Dependent Electron Liquid
  Correlation Energies for Local Spin Density Calculations: A Critical
  Analysis. \emph{Can. J. Phys.} \textbf{1980}, \emph{58}, 1200--1211\relax
\mciteBstWouldAddEndPuncttrue
\mciteSetBstMidEndSepPunct{\mcitedefaultmidpunct}
{\mcitedefaultendpunct}{\mcitedefaultseppunct}\relax
\EndOfBibitem
\bibitem[Miehlich \latin{et~al.}(1989)Miehlich, Savin, Stoll, and
  Preuss]{VRG:miehlich:1989:CPL}
Miehlich,~B.; Savin,~A.; Stoll,~H.; Preuss,~H. Results Obtained with the
  Correlation Energy Density Functionals of Becke and {{Lee}}, {{Yang}} and
  {{Parr}}. \emph{Chem. Phys. Lett.} \textbf{1989}, \emph{157}, 200--206\relax
\mciteBstWouldAddEndPuncttrue
\mciteSetBstMidEndSepPunct{\mcitedefaultmidpunct}
{\mcitedefaultendpunct}{\mcitedefaultseppunct}\relax
\EndOfBibitem
\bibitem[Perdew \latin{et~al.}(1996)Perdew, Burke, and
  Ernzerhof]{VRG:perdew:1996:PRL}
Perdew,~J.~P.; Burke,~K.; Ernzerhof,~M. Generalized Gradient Approximation Made
  Simple. \emph{Phys. Rev. Lett.} \textbf{1996}, \emph{77}, 3865--3868\relax
\mciteBstWouldAddEndPuncttrue
\mciteSetBstMidEndSepPunct{\mcitedefaultmidpunct}
{\mcitedefaultendpunct}{\mcitedefaultseppunct}\relax
\EndOfBibitem
\bibitem[Stephens \latin{et~al.}(1994)Stephens, Devlin, Chabalowski, and
  Frisch]{VRG:stephens:1994:JPC}
Stephens,~P.~J.; Devlin,~F.~J.; Chabalowski,~C.~F.; Frisch,~M.~J. Ab {{Initio
  Calculation}} of {{Vibrational Absorption}} and {{Circular Dichroism Spectra
  Using Density Functional Force Fields}}. \emph{J. Phys. Chem.} \textbf{1994},
  \emph{98}, 11623--11627\relax
\mciteBstWouldAddEndPuncttrue
\mciteSetBstMidEndSepPunct{\mcitedefaultmidpunct}
{\mcitedefaultendpunct}{\mcitedefaultseppunct}\relax
\EndOfBibitem
\bibitem[Adamo and Barone(1999)Adamo, and Barone]{VRG:adamo:1999:JCP}
Adamo,~C.; Barone,~V. Toward Reliable Density Functional Methods without
  Adjustable Parameters: {{The PBE0}} Model. \emph{J. Chem. Phys.}
  \textbf{1999}, \emph{110}, 6158--6170\relax
\mciteBstWouldAddEndPuncttrue
\mciteSetBstMidEndSepPunct{\mcitedefaultmidpunct}
{\mcitedefaultendpunct}{\mcitedefaultseppunct}\relax
\EndOfBibitem
\bibitem[Ernzerhof and Scuseria(1999)Ernzerhof, and
  Scuseria]{VRG:ernzerhof:1999:JCP}
Ernzerhof,~M.; Scuseria,~G.~E. Assessment of the
  {{Perdew}}--{{Burke}}--{{Ernzerhof}} Exchange-Correlation Functional.
  \emph{J. Chem. Phys.} \textbf{1999}, \emph{110}, 5029--5036\relax
\mciteBstWouldAddEndPuncttrue
\mciteSetBstMidEndSepPunct{\mcitedefaultmidpunct}
{\mcitedefaultendpunct}{\mcitedefaultseppunct}\relax
\EndOfBibitem
\bibitem[Zhang and Yang(1998)Zhang, and Yang]{VRG:zhang:1998:PRL}
Zhang,~Y.; Yang,~W. Comment on ``{{Generalized Gradient Approximation Made
  Simple}}''. \emph{Phys. Rev. Lett.} \textbf{1998}, \emph{80}, 890--890\relax
\mciteBstWouldAddEndPuncttrue
\mciteSetBstMidEndSepPunct{\mcitedefaultmidpunct}
{\mcitedefaultendpunct}{\mcitedefaultseppunct}\relax
\EndOfBibitem
\bibitem[Yu \latin{et~al.}(2016)Yu, He, Li, and Truhlar]{VRG:yu:2016:CS}
Yu,~H.~S.; He,~X.; Li,~S.~L.; Truhlar,~D.~G. {{MN15}}: {{A Kohn}}--{{Sham}}
  Global-Hybrid Exchange--Correlation Density Functional with Broad Accuracy
  for Multi-Reference and Single-Reference Systems and Noncovalent
  Interactions. \emph{Chem. Sci.} \textbf{2016}, \emph{7}, 5032--5051\relax
\mciteBstWouldAddEndPuncttrue
\mciteSetBstMidEndSepPunct{\mcitedefaultmidpunct}
{\mcitedefaultendpunct}{\mcitedefaultseppunct}\relax
\EndOfBibitem
\bibitem[Tao \latin{et~al.}(2003)Tao, Perdew, Staroverov, and
  Scuseria]{VRG:tao:2003:PRL}
Tao,~J.; Perdew,~J.~P.; Staroverov,~V.~N.; Scuseria,~G.~E. Climbing the
  {{Density Functional Ladder}}: {{Nonempirical Meta}}--{{Generalized Gradient
  Approximation Designed}} for {{Molecules}} and {{Solids}}. \emph{Phys. Rev.
  Lett.} \textbf{2003}, \emph{91}, 146401\relax
\mciteBstWouldAddEndPuncttrue
\mciteSetBstMidEndSepPunct{\mcitedefaultmidpunct}
{\mcitedefaultendpunct}{\mcitedefaultseppunct}\relax
\EndOfBibitem
\bibitem[Sun \latin{et~al.}(2015)Sun, Ruzsinszky, and Perdew]{VRG:sun:2015:PRL}
Sun,~J.; Ruzsinszky,~A.; Perdew,~J.~P. Strongly {{Constrained}} and
  {{Appropriately Normed Semilocal Density Functional}}. \emph{Phys. Rev.
  Lett.} \textbf{2015}, \emph{115}, 036402\relax
\mciteBstWouldAddEndPuncttrue
\mciteSetBstMidEndSepPunct{\mcitedefaultmidpunct}
{\mcitedefaultendpunct}{\mcitedefaultseppunct}\relax
\EndOfBibitem
\bibitem[Mezei \latin{et~al.}(2018)Mezei, Csonka, and
  K{\'a}llay]{VRG:mezei:2018:JCTC}
Mezei,~P.~D.; Csonka,~G.~I.; K{\'a}llay,~M. Simple {{Modifications}} of the
  {{SCAN Meta-Generalized Gradient Approximation Functional}}. \emph{J. Chem.
  Theory Comput.} \textbf{2018}, \emph{14}, 2469--2479\relax
\mciteBstWouldAddEndPuncttrue
\mciteSetBstMidEndSepPunct{\mcitedefaultmidpunct}
{\mcitedefaultendpunct}{\mcitedefaultseppunct}\relax
\EndOfBibitem
\bibitem[Yanai \latin{et~al.}(2004)Yanai, Tew, and Handy]{VRG:yanai:2004:CPL}
Yanai,~T.; Tew,~D.~P.; Handy,~N.~C. A New Hybrid Exchange--Correlation
  Functional Using the {{Coulomb-attenuating}} Method ({{CAM-B3LYP}}).
  \emph{Chem. Phys. Lett.} \textbf{2004}, \emph{393}, 51--57\relax
\mciteBstWouldAddEndPuncttrue
\mciteSetBstMidEndSepPunct{\mcitedefaultmidpunct}
{\mcitedefaultendpunct}{\mcitedefaultseppunct}\relax
\EndOfBibitem
\bibitem[Chai and {Head-Gordon}(2008)Chai, and
  {Head-Gordon}]{VRG:chai:2008:JCP}
Chai,~J.-D.; {Head-Gordon},~M. Systematic Optimization of Long-Range Corrected
  Hybrid Density Functionals. \emph{J. Chem. Phys.} \textbf{2008}, \emph{128},
  084106\relax
\mciteBstWouldAddEndPuncttrue
\mciteSetBstMidEndSepPunct{\mcitedefaultmidpunct}
{\mcitedefaultendpunct}{\mcitedefaultseppunct}\relax
\EndOfBibitem
\bibitem[Henderson \latin{et~al.}(2008)Henderson, Janesko, and
  Scuseria]{VRG:henderson:2008:JCP}
Henderson,~T.~M.; Janesko,~B.~G.; Scuseria,~G.~E. Generalized Gradient
  Approximation Model Exchange Holes for Range-Separated Hybrids. \emph{J.
  Chem. Phys.} \textbf{2008}, \emph{128}, 194105\relax
\mciteBstWouldAddEndPuncttrue
\mciteSetBstMidEndSepPunct{\mcitedefaultmidpunct}
{\mcitedefaultendpunct}{\mcitedefaultseppunct}\relax
\EndOfBibitem
\bibitem[Weintraub \latin{et~al.}(2009)Weintraub, Henderson, and
  Scuseria]{VRG:weintraub:2009:JCTC}
Weintraub,~E.; Henderson,~T.~M.; Scuseria,~G.~E. Long-{{Range-Corrected Hybrids
  Based}} on a {{New Model Exchange Hole}}. \emph{J. Chem. Theory Comput.}
  \textbf{2009}, \emph{5}, 754--762\relax
\mciteBstWouldAddEndPuncttrue
\mciteSetBstMidEndSepPunct{\mcitedefaultmidpunct}
{\mcitedefaultendpunct}{\mcitedefaultseppunct}\relax
\EndOfBibitem
\bibitem[Ditchfield \latin{et~al.}(1971)Ditchfield, Hehre, and
  Pople]{VRG:ditchfield:1971:JCP}
Ditchfield,~R.; Hehre,~W.~J.; Pople,~J.~A. Self-{{Consistent
  Molecular}}-{{Orbital Methods}}. {{IX}}. {{An Extended Gaussian}}-{{Type
  Basis}} for {{Molecular}}-{{Orbital Studies}} of {{Organic Molecules}}.
  \emph{J. Chem. Phys.} \textbf{1971}, \emph{54}, 724--728\relax
\mciteBstWouldAddEndPuncttrue
\mciteSetBstMidEndSepPunct{\mcitedefaultmidpunct}
{\mcitedefaultendpunct}{\mcitedefaultseppunct}\relax
\EndOfBibitem
\bibitem[Hehre \latin{et~al.}(1972)Hehre, Ditchfield, and
  Pople]{VRG:hehre:1972:JCP}
Hehre,~W.~J.; Ditchfield,~R.; Pople,~J.~A. Self---{{Consistent}} Molecular
  Orbital Methods. {{XII}}. {{Further}} Extensions of {{Gaussian}}---{{Type}}
  Basis Sets for Use in Molecular Orbital Studies of Organic Molecules.
  \emph{J. Chem. Phys.} \textbf{1972}, \emph{56}, 2257--2261\relax
\mciteBstWouldAddEndPuncttrue
\mciteSetBstMidEndSepPunct{\mcitedefaultmidpunct}
{\mcitedefaultendpunct}{\mcitedefaultseppunct}\relax
\EndOfBibitem
\bibitem[Hariharan and Pople(1973)Hariharan, and Pople]{VRG:hariharan:1973:TCA}
Hariharan,~P.~C.; Pople,~J.~A. The Influence of Polarization Functions on
  Molecular Orbital Hydrogenation Energies. \emph{Theoret. Chim. Acta}
  \textbf{1973}, \emph{28}, 213--222\relax
\mciteBstWouldAddEndPuncttrue
\mciteSetBstMidEndSepPunct{\mcitedefaultmidpunct}
{\mcitedefaultendpunct}{\mcitedefaultseppunct}\relax
\EndOfBibitem
\bibitem[Francl \latin{et~al.}(1982)Francl, Pietro, Hehre, Binkley, Gordon,
  DeFrees, and Pople]{VRG:francl:1982:JCP}
Francl,~M.~M.; Pietro,~W.~J.; Hehre,~W.~J.; Binkley,~J.~S.; Gordon,~M.~S.;
  DeFrees,~D.~J.; Pople,~J.~A. Self-consistent Molecular Orbital Methods.
  {{XXIII}}. {{A}} Polarization-type Basis Set for Second-row Elements.
  \emph{J. Chem. Phys.} \textbf{1982}, \emph{77}, 3654--3665\relax
\mciteBstWouldAddEndPuncttrue
\mciteSetBstMidEndSepPunct{\mcitedefaultmidpunct}
{\mcitedefaultendpunct}{\mcitedefaultseppunct}\relax
\EndOfBibitem
\bibitem[Gordon \latin{et~al.}(1982)Gordon, Binkley, Pople, Pietro, and
  Hehre]{VRG:gordon:1982:JACS}
Gordon,~M.~S.; Binkley,~J.~S.; Pople,~J.~A.; Pietro,~W.~J.; Hehre,~W.~J.
  Self-Consistent Molecular-Orbital Methods. 22. {{Small}} Split-Valence Basis
  Sets for Second-Row Elements. \emph{J. Am. Chem. Soc.} \textbf{1982},
  \emph{104}, 2797--2803\relax
\mciteBstWouldAddEndPuncttrue
\mciteSetBstMidEndSepPunct{\mcitedefaultmidpunct}
{\mcitedefaultendpunct}{\mcitedefaultseppunct}\relax
\EndOfBibitem
\bibitem[Weigend(2006)]{VRG:weigend:2006:PCCP}
Weigend,~F. Accurate {{Coulomb-fitting}} Basis Sets for {{H}} to {{Rn}}.
  \emph{Phys. Chem. Chem. Phys.} \textbf{2006}, \emph{8}, 1057\relax
\mciteBstWouldAddEndPuncttrue
\mciteSetBstMidEndSepPunct{\mcitedefaultmidpunct}
{\mcitedefaultendpunct}{\mcitedefaultseppunct}\relax
\EndOfBibitem
\bibitem[Berman(2000)]{VRG:berman:2000:NAR}
Berman,~H.~M. The {{Protein Data Bank}}. \emph{Nucleic Acids Res.}
  \textbf{2000}, \emph{28}, 235--242\relax
\mciteBstWouldAddEndPuncttrue
\mciteSetBstMidEndSepPunct{\mcitedefaultmidpunct}
{\mcitedefaultendpunct}{\mcitedefaultseppunct}\relax
\EndOfBibitem
\bibitem[Smith \latin{et~al.}(2020)Smith, Burns, Simmonett, Parrish, Schieber,
  Galvelis, Kraus, Kruse, Di~Remigio, Alenaizan, James, Lehtola, Misiewicz,
  Scheurer, Shaw, Schriber, Xie, Glick, Sirianni, O'Brien, Waldrop, Kumar,
  Hohenstein, Pritchard, Brooks, SchaeferIII, Sokolov, Patkowski, DePrinceIII,
  Bozkaya, King, Evangelista, Turney, Crawford, and
  Sherrill]{VRG:smith:2020:JCP}
Smith,~D. G.~A.; Burns,~L.~A.; Simmonett,~A.~C.; Parrish,~R.~M.;
  Schieber,~M.~C.; Galvelis,~R.; Kraus,~P.; Kruse,~H.; Di~Remigio,~R.;
  Alenaizan,~A.; James,~A.~M.; Lehtola,~S.; Misiewicz,~J.~P.; Scheurer,~M.;
  Shaw,~R.~A.; Schriber,~J.~B.; Xie,~Y.; Glick,~Z.~L.; Sirianni,~D.~A.;
  O'Brien,~J.~S.; Waldrop,~J.~M.; Kumar,~A.; Hohenstein,~E.~G.;
  Pritchard,~B.~P.; Brooks,~B.~R.; SchaeferIII,~H.~F.; Sokolov,~A.~Y.;
  Patkowski,~K.; DePrinceIII,~A.~E.; Bozkaya,~U.; King,~R.~A.;
  Evangelista,~F.~A.; Turney,~J.~M.; Crawford,~T.~D.; Sherrill,~C.~D. {{PSI4}}
  1.4: {{Open-source}} Software for High-Throughput Quantum Chemistry. \emph{J.
  Chem. Phys.} \textbf{2020}, \emph{152}, 184108\relax
\mciteBstWouldAddEndPuncttrue
\mciteSetBstMidEndSepPunct{\mcitedefaultmidpunct}
{\mcitedefaultendpunct}{\mcitedefaultseppunct}\relax
\EndOfBibitem
\bibitem[Janak(1978)]{VRG:janak:1978:PRB}
Janak,~J.~F. Proof That {$\partial$} {{E}} {$\partial$} n i = {$\varepsilon$}
  in Density-Functional Theory. \emph{Phys. Rev. B} \textbf{1978}, \emph{18},
  7165--7168\relax
\mciteBstWouldAddEndPuncttrue
\mciteSetBstMidEndSepPunct{\mcitedefaultmidpunct}
{\mcitedefaultendpunct}{\mcitedefaultseppunct}\relax
\EndOfBibitem
\bibitem[Johnson \latin{et~al.}(2013)Johnson, Salamone, Bietti, and
  DiLabio]{VRG:johnson:2013:JPCA}
Johnson,~E.~R.; Salamone,~M.; Bietti,~M.; DiLabio,~G.~A. Modeling {{Noncovalent
  Radical}}--{{Molecule Interactions Using Conventional Density-Functional
  Theory}}: {{Beware Erroneous Charge Transfer}}. \emph{J. Phys. Chem. A}
  \textbf{2013}, \emph{117}, 947--952\relax
\mciteBstWouldAddEndPuncttrue
\mciteSetBstMidEndSepPunct{\mcitedefaultmidpunct}
{\mcitedefaultendpunct}{\mcitedefaultseppunct}\relax
\EndOfBibitem
\bibitem[Sini \latin{et~al.}(2011)Sini, Sears, and
  Br{\'e}das]{VRG:sini:2011:JCTC}
Sini,~G.; Sears,~J.~S.; Br{\'e}das,~J.-L. Evaluating the {{Performance}} of
  {{DFT Functionals}} in {{Assessing}} the {{Interaction Energy}} and
  {{Ground-State Charge Transfer}} of {{Donor}}/{{Acceptor Complexes}}:
  {{Tetrathiafulvalene}}-{{Tetracyanoquinodimethane}} ({{TTF}}-{{TCNQ}}) as a
  {{Model Case}}. \emph{J. Chem. Theory Comput.} \textbf{2011}, \emph{7},
  602--609\relax
\mciteBstWouldAddEndPuncttrue
\mciteSetBstMidEndSepPunct{\mcitedefaultmidpunct}
{\mcitedefaultendpunct}{\mcitedefaultseppunct}\relax
\EndOfBibitem
\bibitem[Prasad \latin{et~al.}(2019)Prasad, {Otero-de-la-Roza}, and
  DiLabio]{VRG:prasad:2019:SD}
Prasad,~V.~K.; {Otero-de-la-Roza},~A.; DiLabio,~G.~A. {{PEPCONF}}, a Diverse
  Data Set of Peptide Conformational Energies. \emph{Scientific Data}
  \textbf{2019}, \emph{6}, 180310\relax
\mciteBstWouldAddEndPuncttrue
\mciteSetBstMidEndSepPunct{\mcitedefaultmidpunct}
{\mcitedefaultendpunct}{\mcitedefaultseppunct}\relax
\EndOfBibitem
\bibitem[Culka \latin{et~al.}(2021)Culka, Kalvoda, Gutten, and
  Rulíšek]{doi:10.1021/acs.jpcb.0c09251}
Culka,~M.; Kalvoda,~T.; Gutten,~O.; Rulíšek,~L. Mapping Conformational Space
  of All 8000 Tripeptides by Quantum Chemical Methods: What Strain Is
  Affordable within Folded Protein Chains? \emph{J. Phys. Chem. B}
  \textbf{2021}, \emph{125}, 58--69, PMID: 33393778\relax
\mciteBstWouldAddEndPuncttrue
\mciteSetBstMidEndSepPunct{\mcitedefaultmidpunct}
{\mcitedefaultendpunct}{\mcitedefaultseppunct}\relax
\EndOfBibitem
\end{mcitethebibliography}

\end{document}

\section{Structures of Systems from the PDB}
            Chemical structures of the 12 systems investigated in section 3.1 are presented here.

        \begin{figure}
        \centering
        \includegraphics[width=6.5in]{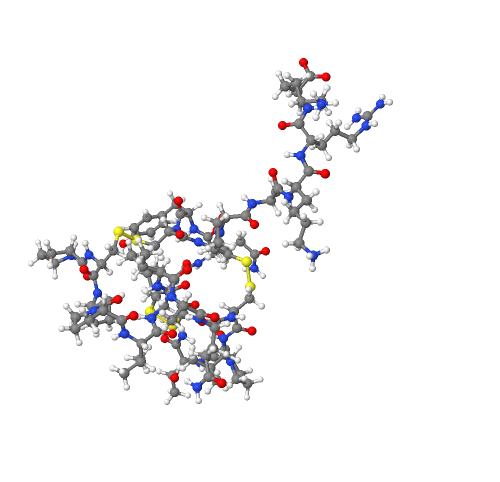}
        \caption{1SP7}
        \label{fig:xyz_1SP7}
        \end{figure}

        \begin{figure}
        \centering
        \includegraphics[width=6.5in]{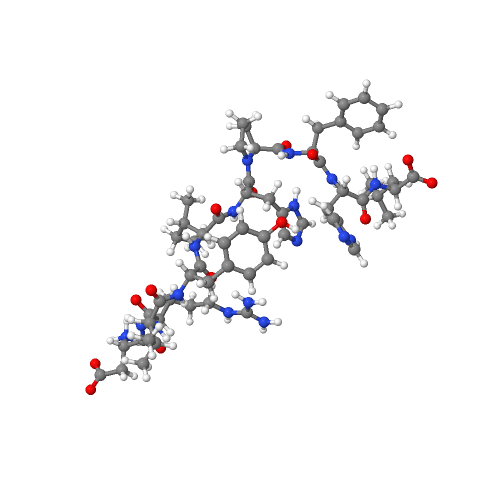}
        \caption{1N9U}
        \label{fig:xyz_1N9U}
        \end{figure}

        \begin{figure}
        \centering
        \includegraphics[width=6.5in]{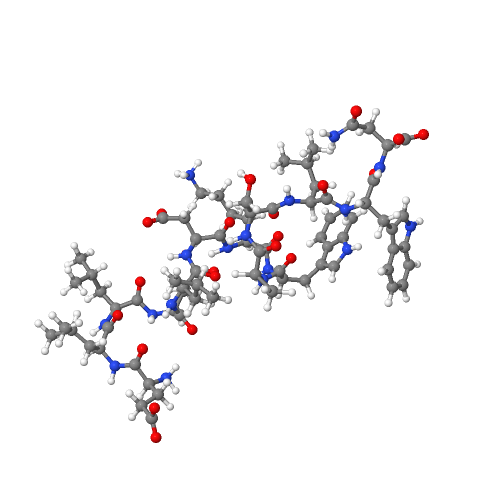}
        \caption{1MZI}
        \label{fig:xyz_1MZI}
        \end{figure}

        \begin{figure}
        \centering
        \includegraphics[width=6.5in]{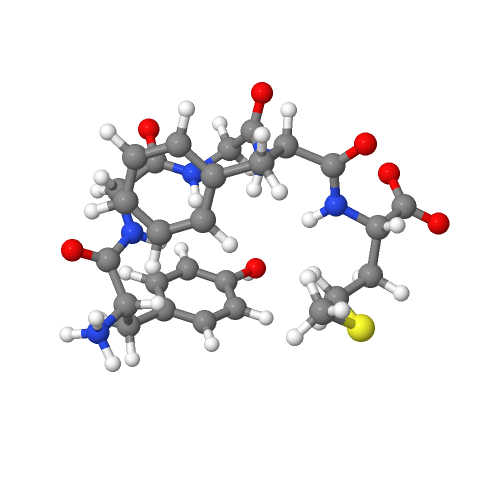}
        \caption{1PLW}
        \label{fig:xyz_1PLW}
        \end{figure}

        \begin{figure}
        \centering
        \includegraphics[width=6.5in]{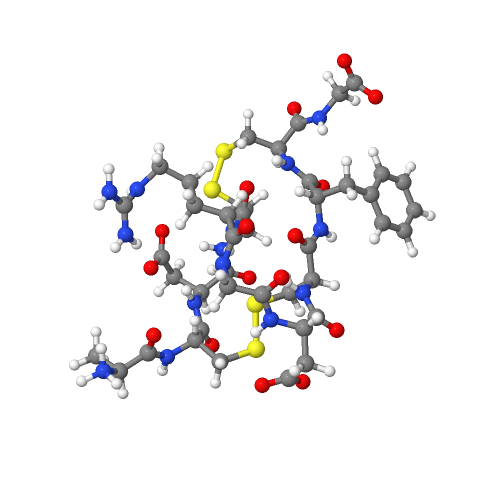}
        \caption{1FUL}
        \label{fig:xyz_1FUL}
        \end{figure}
            
        \begin{figure}
        \centering
        \includegraphics[width=6.5in]{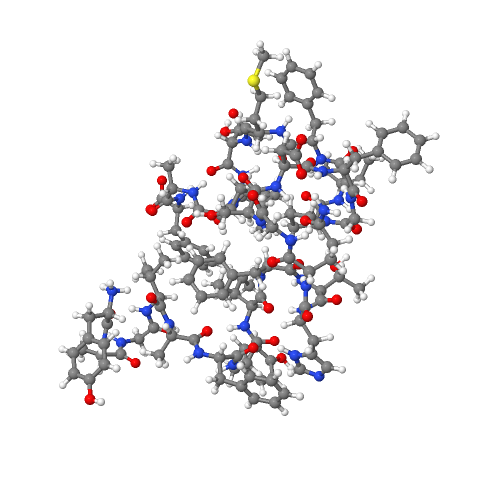}
        \caption{1EDW}
        \label{fig:xyz_1EDW}
        \end{figure}
        
        \begin{figure}
        \centering
        \includegraphics[width=6.5in]{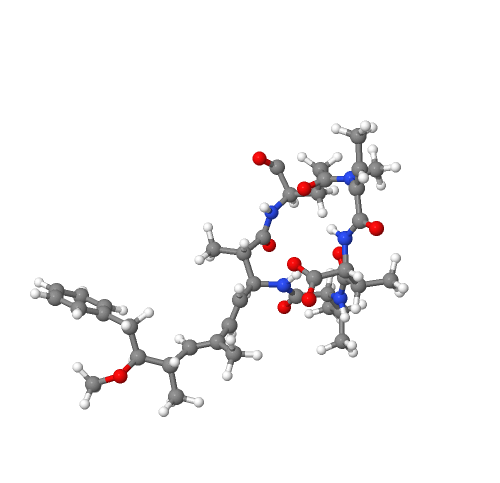}
        \caption{1EVC}
        \label{fig:xyz_1EVC}
        \end{figure}

        \begin{figure}
        \centering
        \includegraphics[width=6.5in]{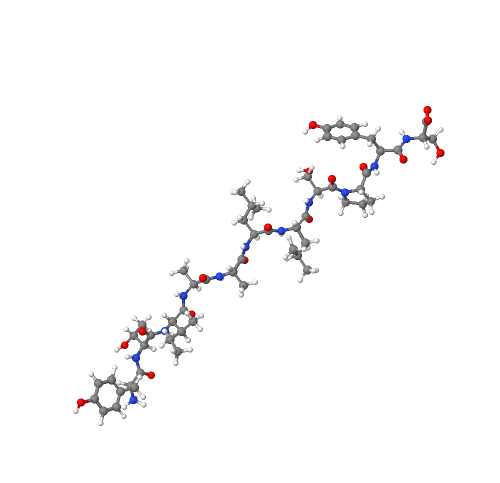}
        \caption{1RVS}
        \label{fig:xyz_1RVS}
        \end{figure}

        \begin{figure}
        \centering
        \includegraphics[width=6.5in]{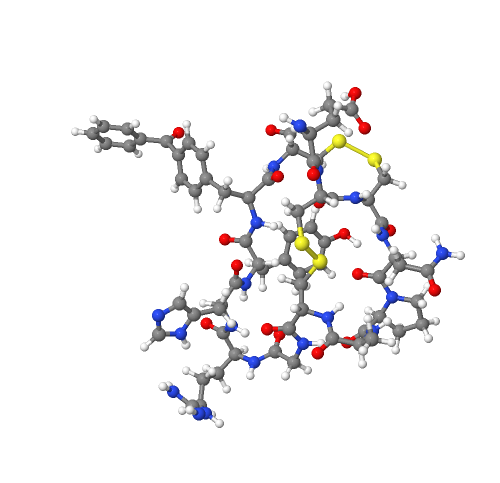}
        \caption{2FR9}
        \label{fig:xyz_2FR9}
        \end{figure}
        
        \begin{figure}
        \centering
        \includegraphics[width=6.5in]{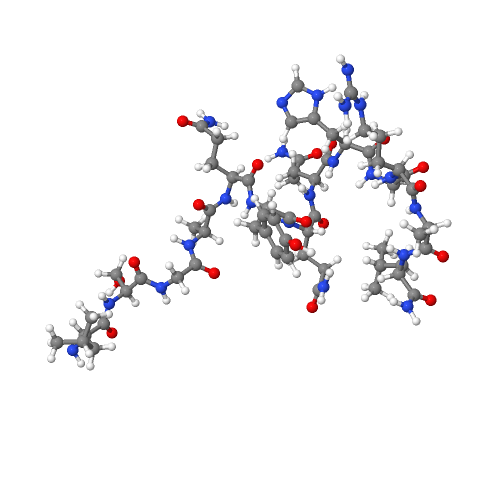}
        \caption{2JSI}
        \label{fig:xyz_2JSI}
        \end{figure}

        \begin{figure}
        \centering
        \includegraphics[width=6.5in]{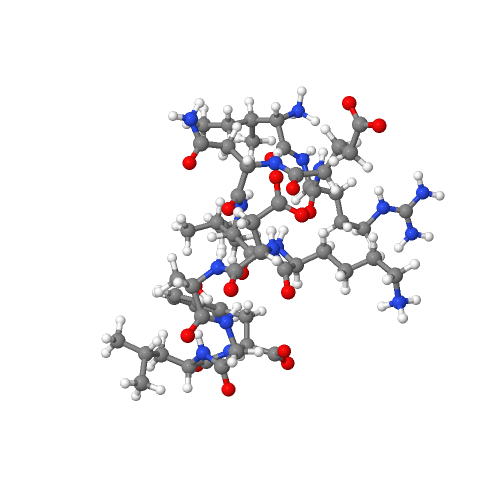}
        \caption{1LVZ}
        \label{fig:xyz_1LVZ}
        \end{figure}
        
        \begin{figure}
        \centering
        \includegraphics[width=6.5in]{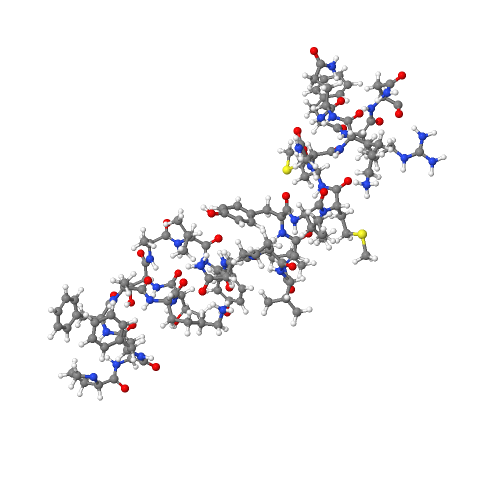}
        \caption{1FDF}
        \label{fig:xyz_1FDF}
        \end{figure}

\section{Natural Deformation Orbitals: Systems from the PDB using LDA}

Images of the HF-LDA natural deformation orbitals with magnitudes of deformation charges greater than 0.2 (``Frontier Natural Deformation Orbitals'', or FNDO) are presented in this section, along with the Hartree-Fock HOMO and LUMO.

        \begin{figure}
        \centering
        \includegraphics[width=7.0in]{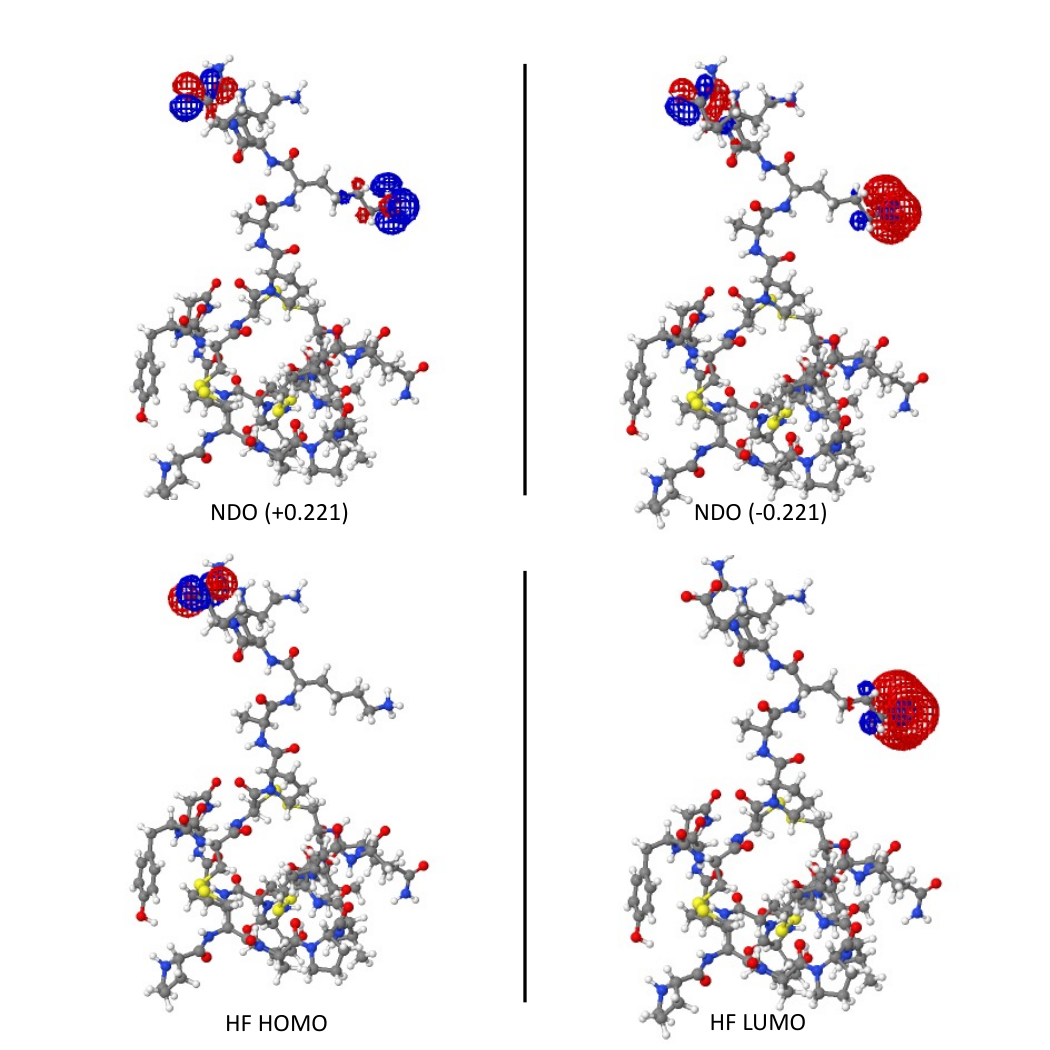}
        \caption{1SP7: HF-LDA FNDOs, juxtaposed with the Hartree-Fock HOMO and LUMO.}
        \label{fig:1SP7_nd_vs_hf}
        \end{figure}

        \begin{figure}
        \centering
        \includegraphics[width=7.0in]{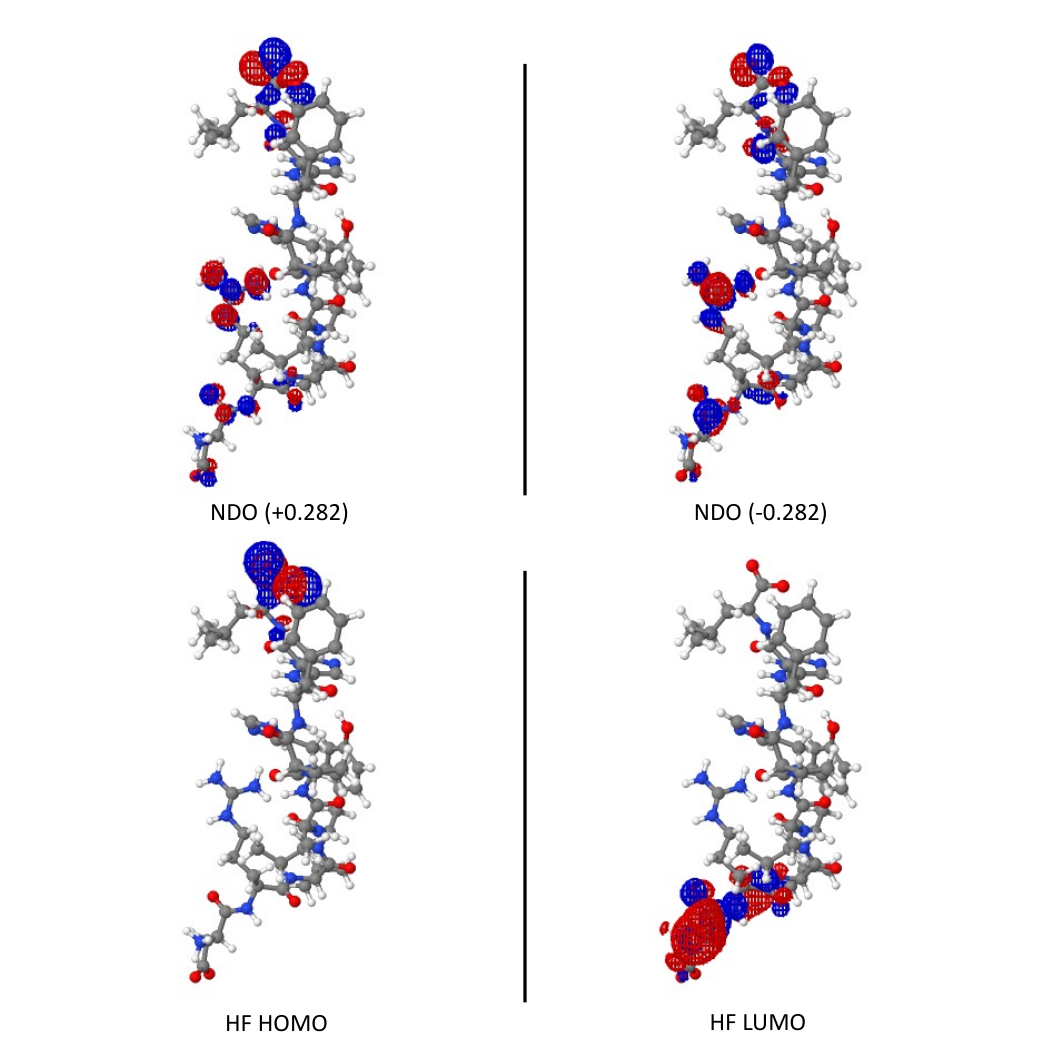}
        \caption{1N9U: HF-LDA FNDOs, juxtaposed with the Hartree-Fock HOMO and LUMO.}
        \label{fig:1N9U_nd_vs_hf}
        \end{figure}

        \begin{figure}
        \centering
        \includegraphics[width=7.0in]{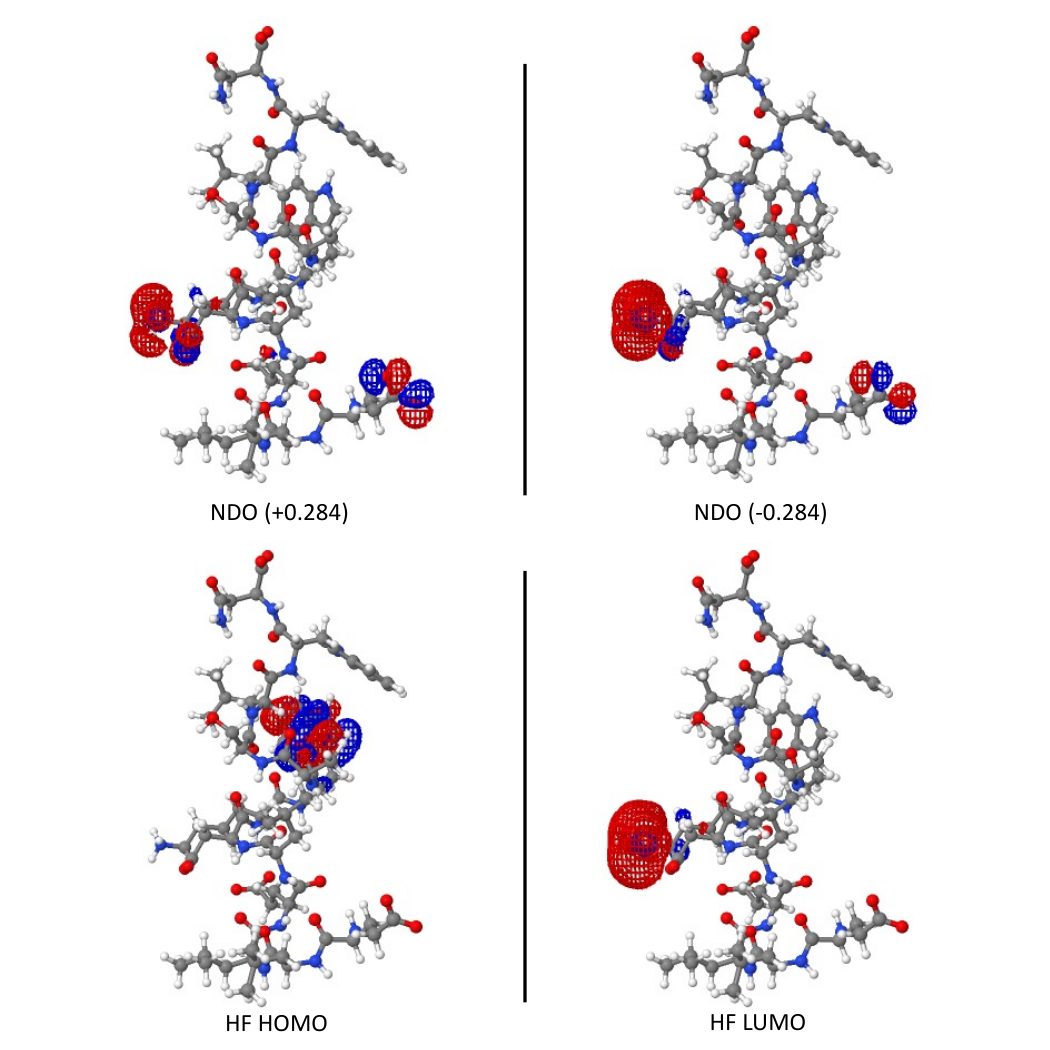}
        \caption{1MZI: HF-LDA FNDOs, juxtaposed with the Hartree-Fock HOMO and LUMO.}
        \label{fig:1MZI_nd_vs_hf}
        \end{figure}

        \begin{figure}
        \centering
        \includegraphics[width=7.0in]{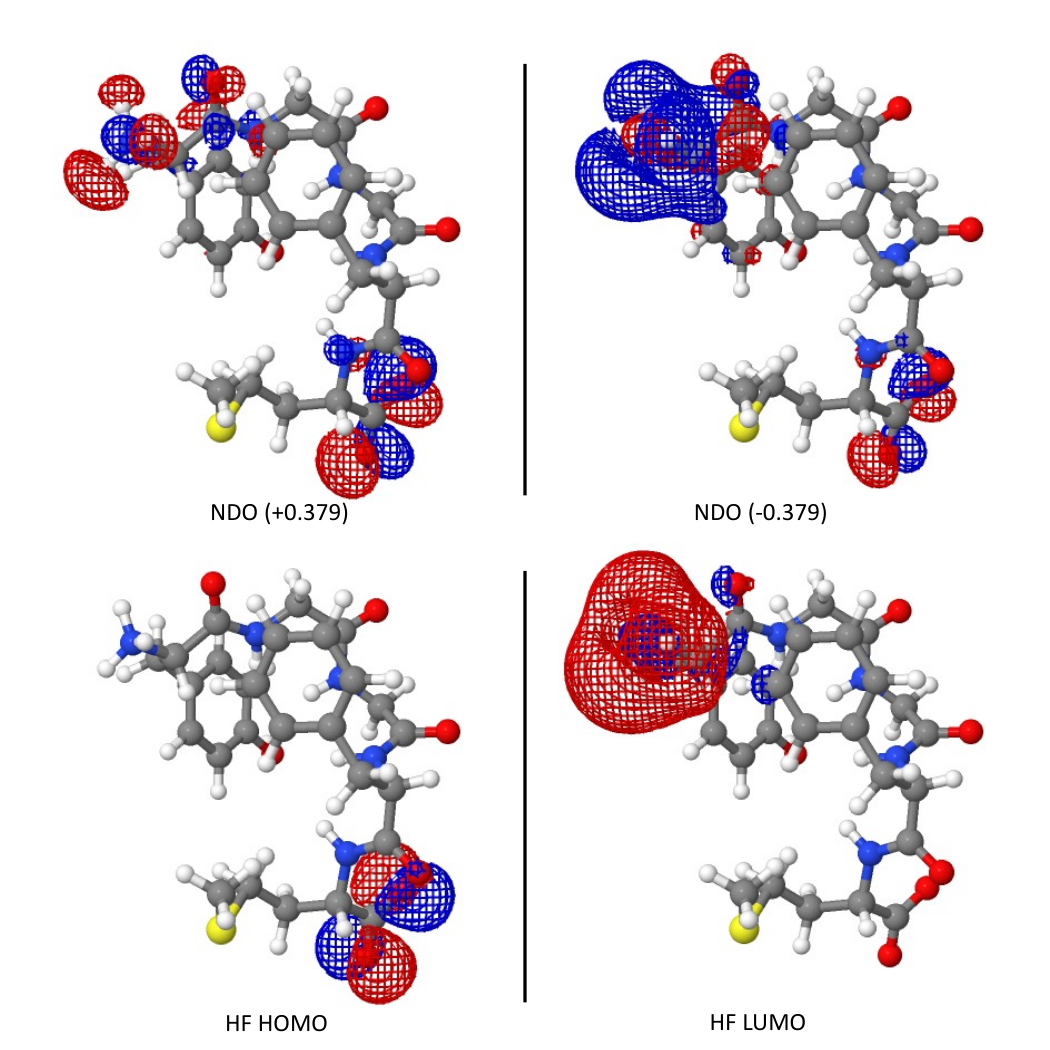}
        \caption{1PLW: HF-LDA FNDOs, juxtaposed with the Hartree-Fock HOMO and LUMO.}
        \label{fig:1PLW_nd_vs_hf}
        \end{figure}

        \begin{figure}
        \centering
        \includegraphics[width=7.0in]{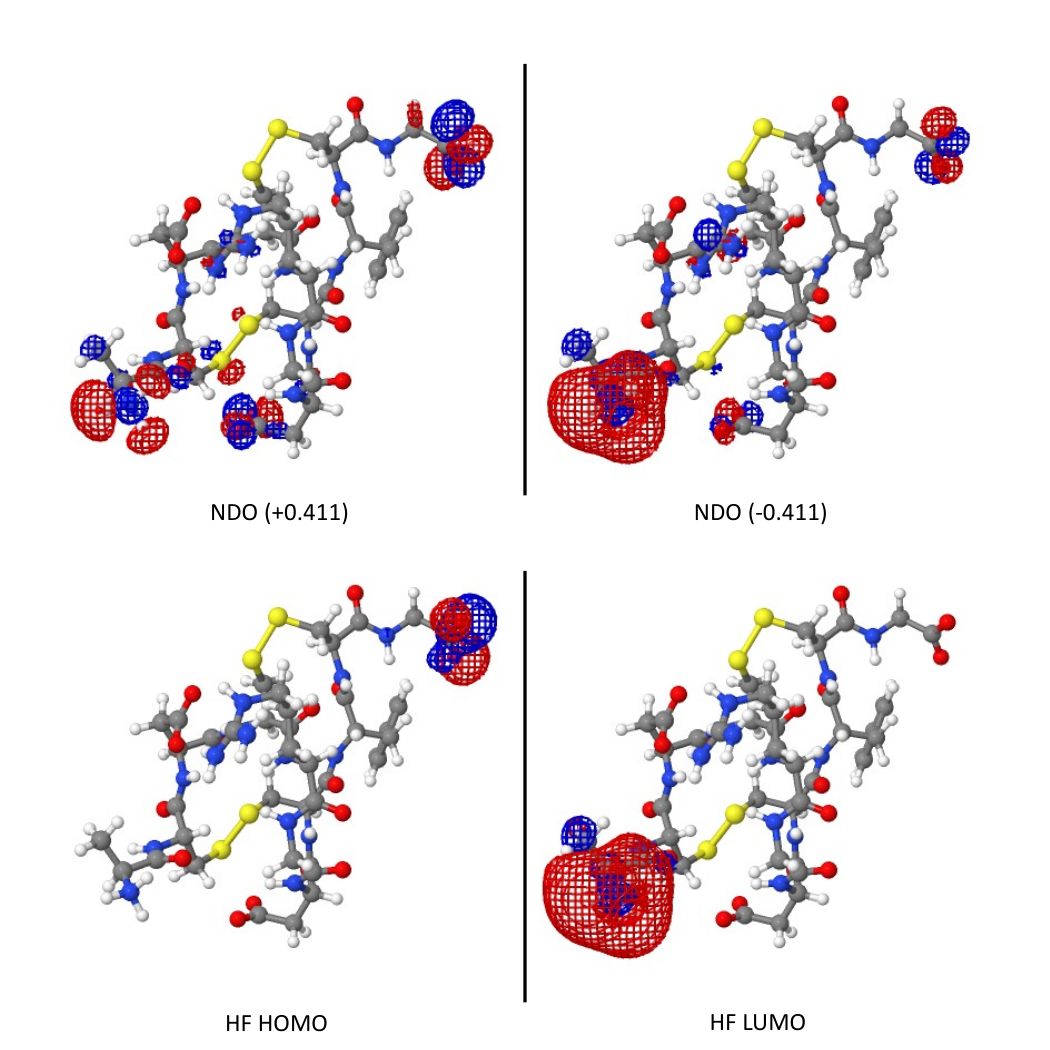}
        \caption{1FUL: HF-LDA FNDOs, juxtaposed with the Hartree-Fock HOMO and LUMO.}
        \label{fig:1FUL_nd_vs_hf}
        \end{figure}

        \begin{figure}
        \centering
        \includegraphics[width=7.0in]{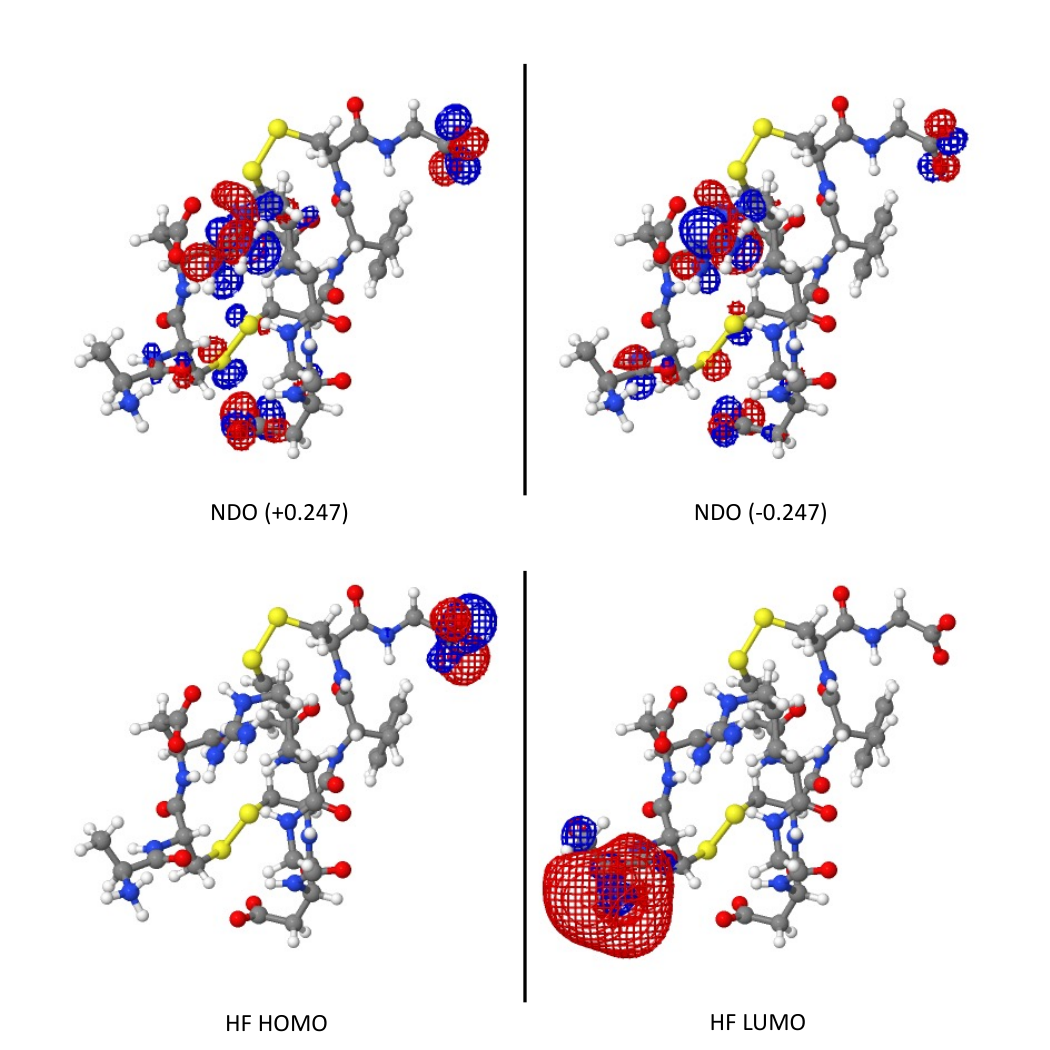}
        \caption{1FUL: HF-LDA FNDOs, juxtaposed with the Hartree-Fock HOMO and LUMO.}
        \label{fig:1FUL_nd_vs_hf_2nd}
        \end{figure}

        \begin{figure}
        \centering
        \includegraphics[width=7.0in]{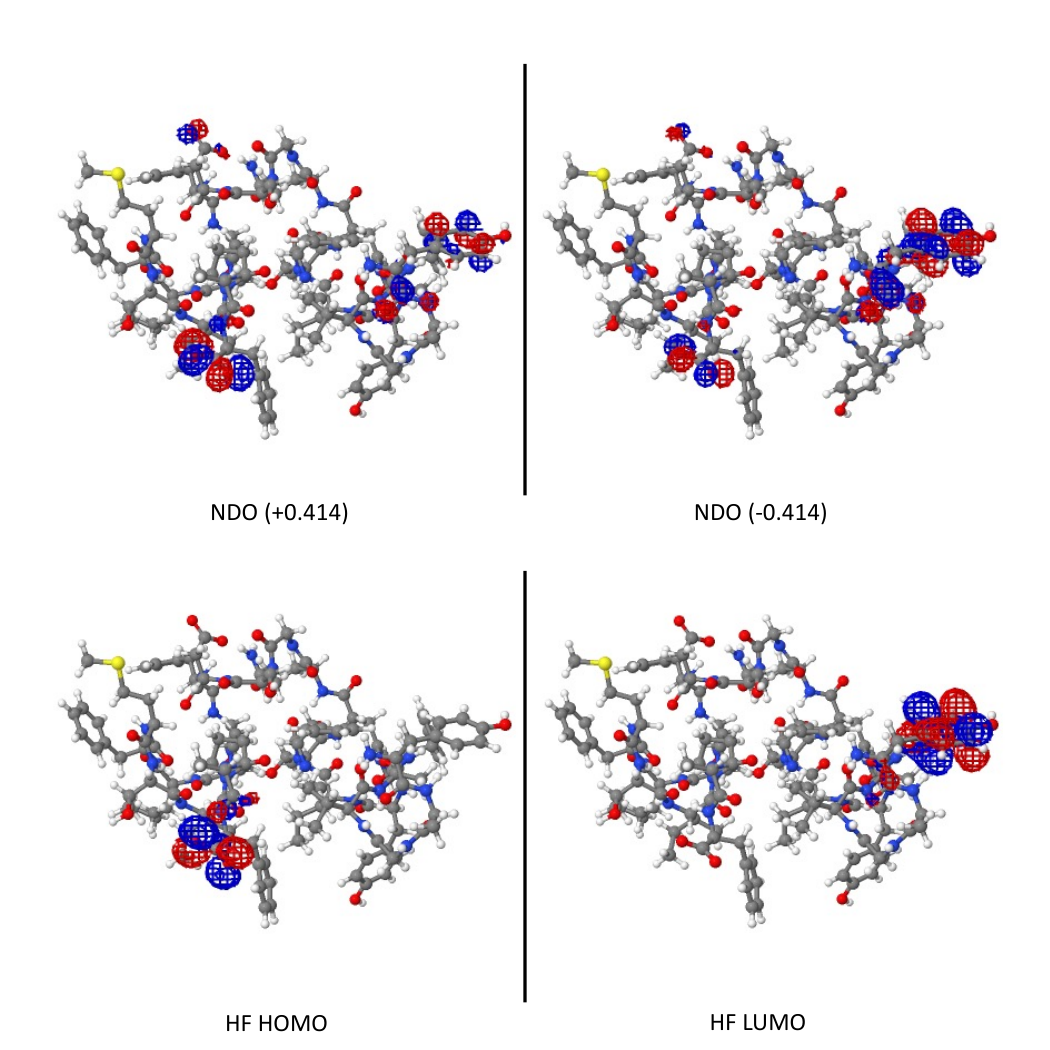}
        \caption{1EDW: HF-LDA FNDOs, juxtaposed with the Hartree-Fock HOMO and LUMO.}
        \label{fig:1EDW_nd_vs_hf}
        \end{figure}

        \begin{figure}
        \centering
        \includegraphics[width=7.0in]{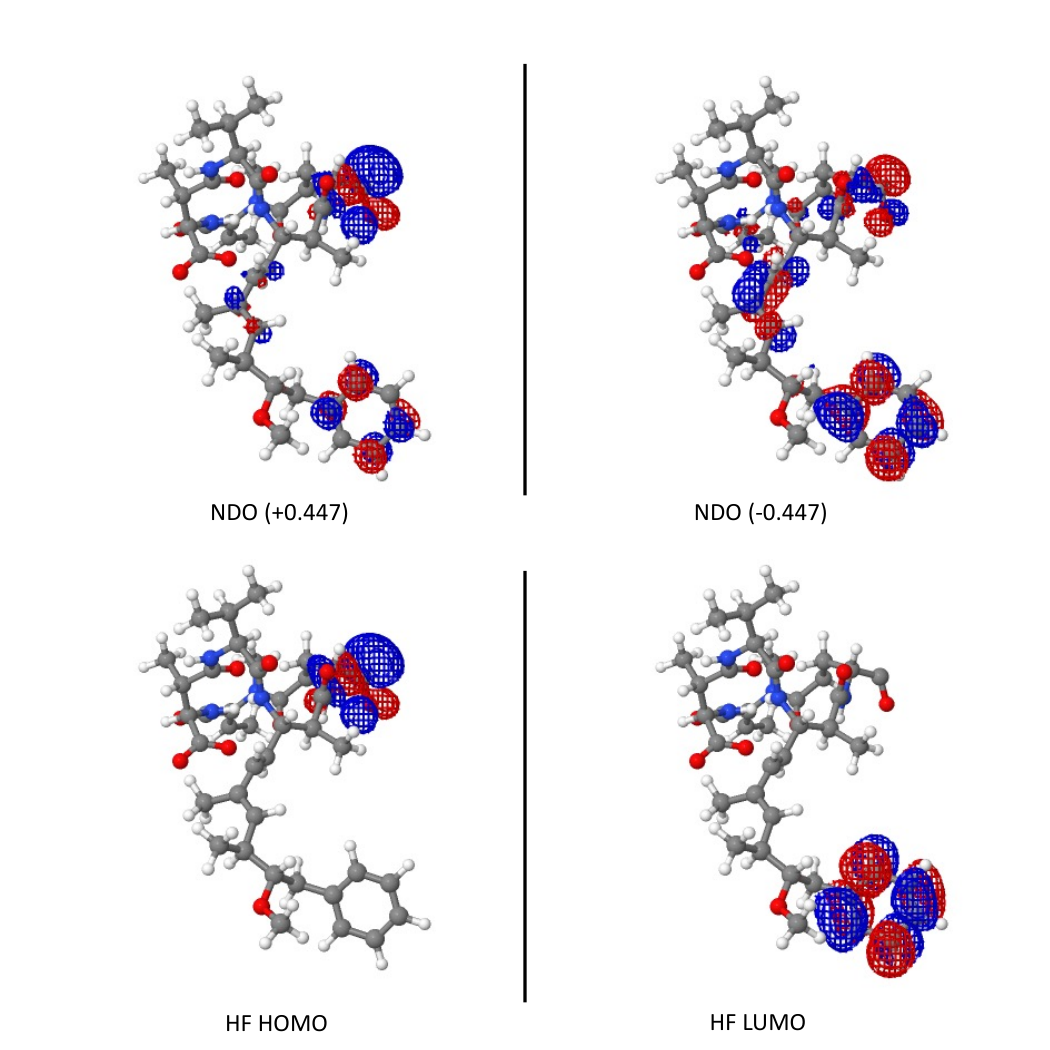}
        \caption{1EVC: HF-LDA FNDOs, juxtaposed with the Hartree-Fock HOMO and LUMO.}
        \label{fig:1EVC_nd_vs_hf}
        \end{figure}

        \begin{figure}
        \centering
        \includegraphics[width=7.0in]{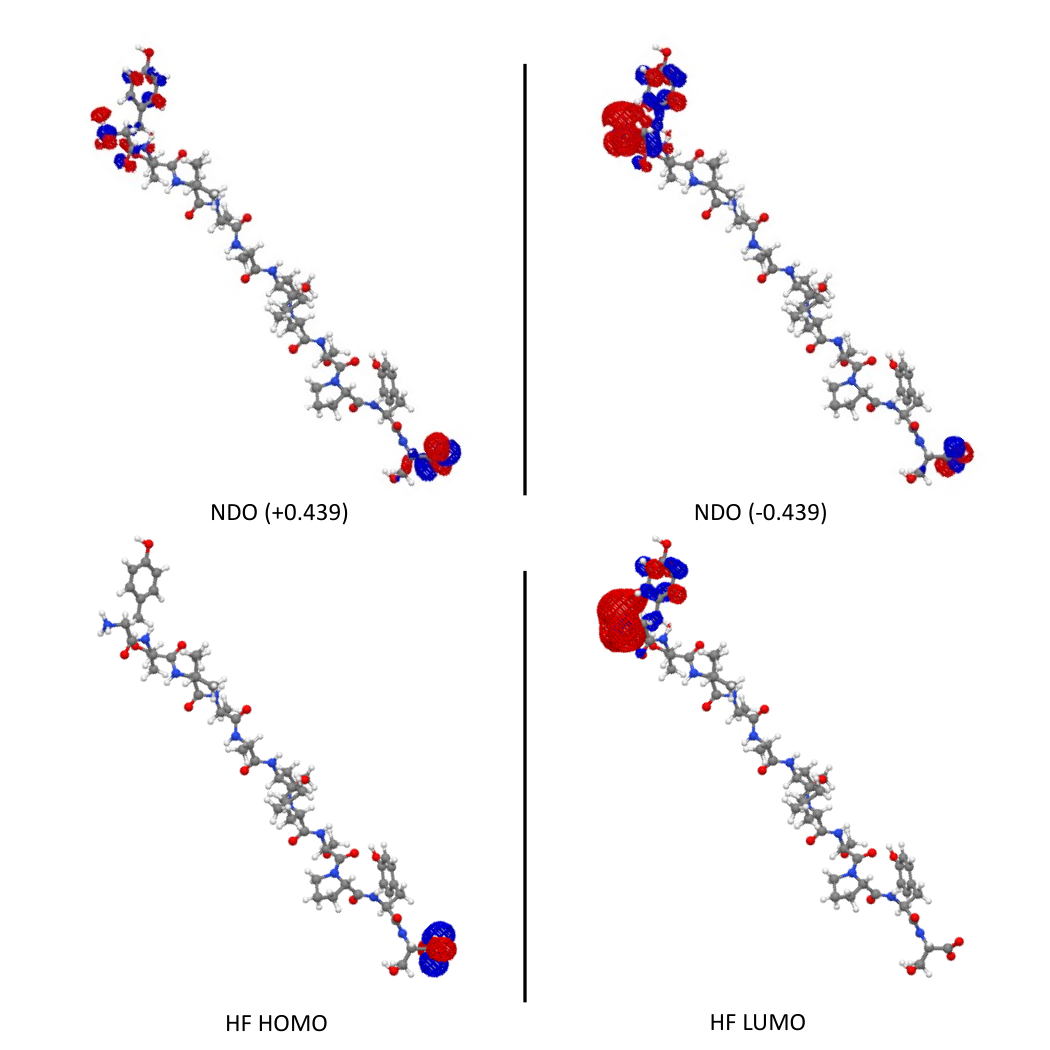}
        \caption{1RVS: HF-LDA FNDOs, juxtaposed with the Hartree-Fock HOMO and LUMO.}
        \label{fig:1RVS_nd_vs_hf}
        \end{figure}

        \begin{figure}
        \centering
        \includegraphics[width=7.0in]{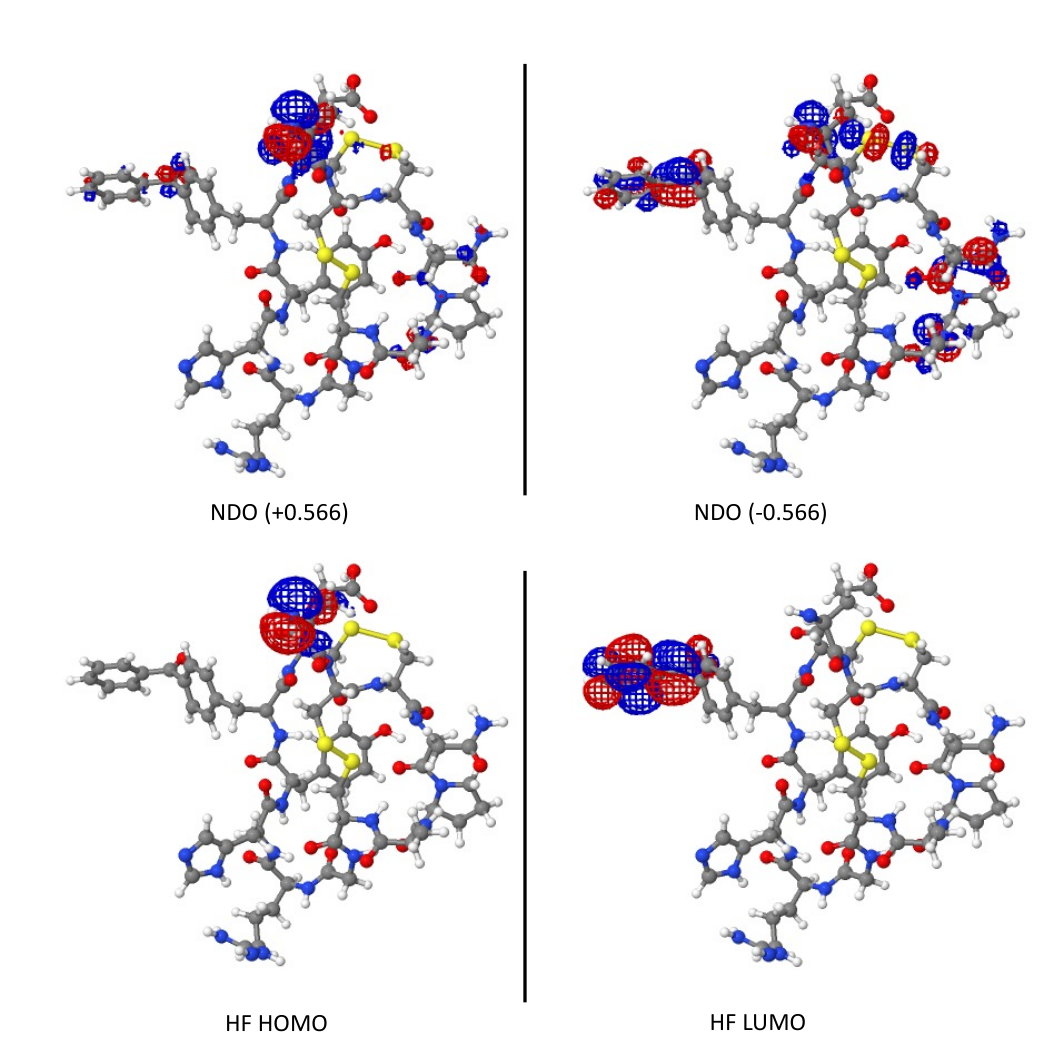}
        \caption{2FR9: HF-LDA FNDOs, juxtaposed with the Hartree-Fock HOMO and LUMO.}
        \label{fig:2FR9_nd_vs_hf}
        \end{figure}

        \begin{figure}
        \centering
        \includegraphics[width=7.0in]{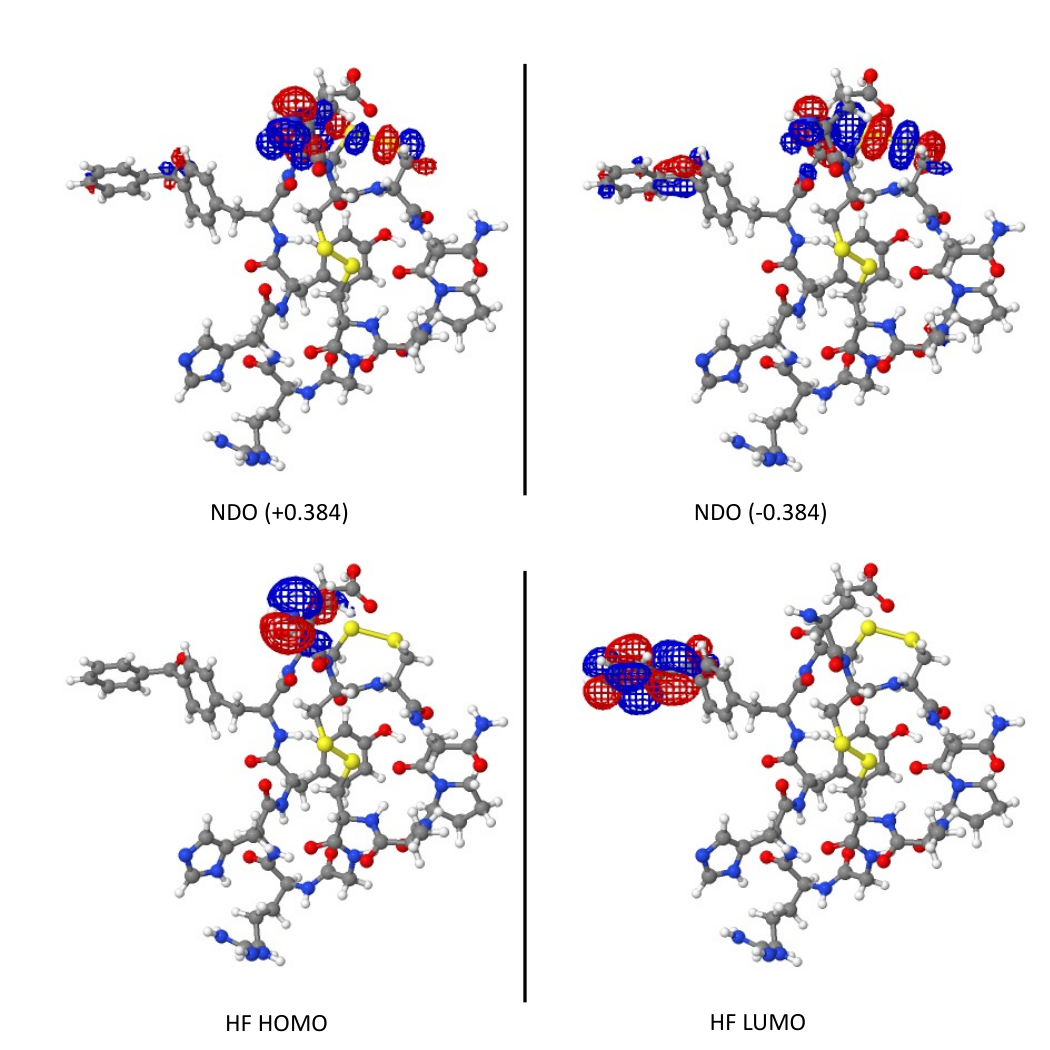}
        \caption{2FR9: HF-LDA FNDOs, juxtaposed with the Hartree-Fock HOMO and LUMO.}
        \label{fig:2FR9_nd_vs_hf_2nd}
        \end{figure}

        \begin{figure}
        \centering
        \includegraphics[width=7.0in]{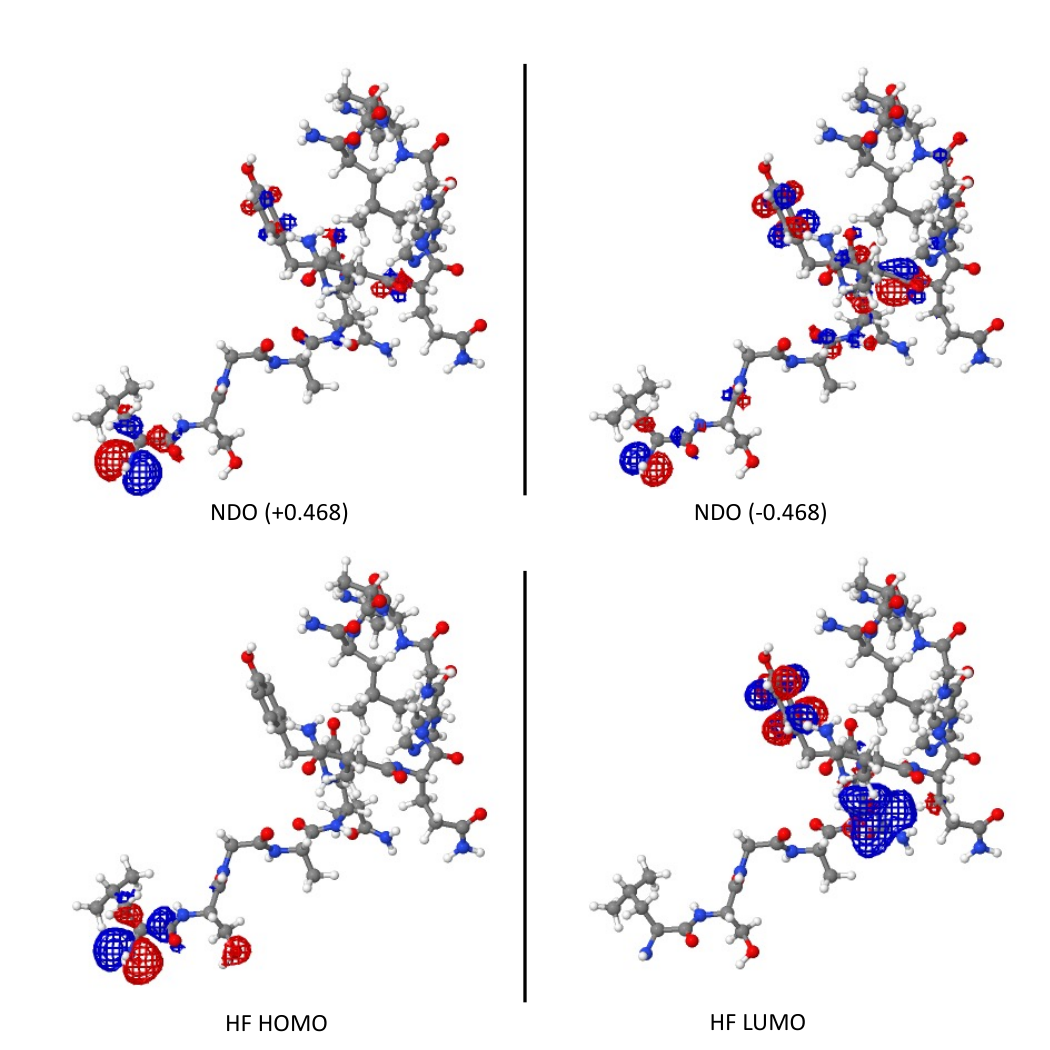}
        \caption{2JSI: HF-LDA FNDOs, juxtaposed with the Hartree-Fock HOMO and LUMO.}
        \label{fig:2JSI_nd_vs_hf}
        \end{figure}

        \begin{figure}
        \centering
        \includegraphics[width=7.0in]{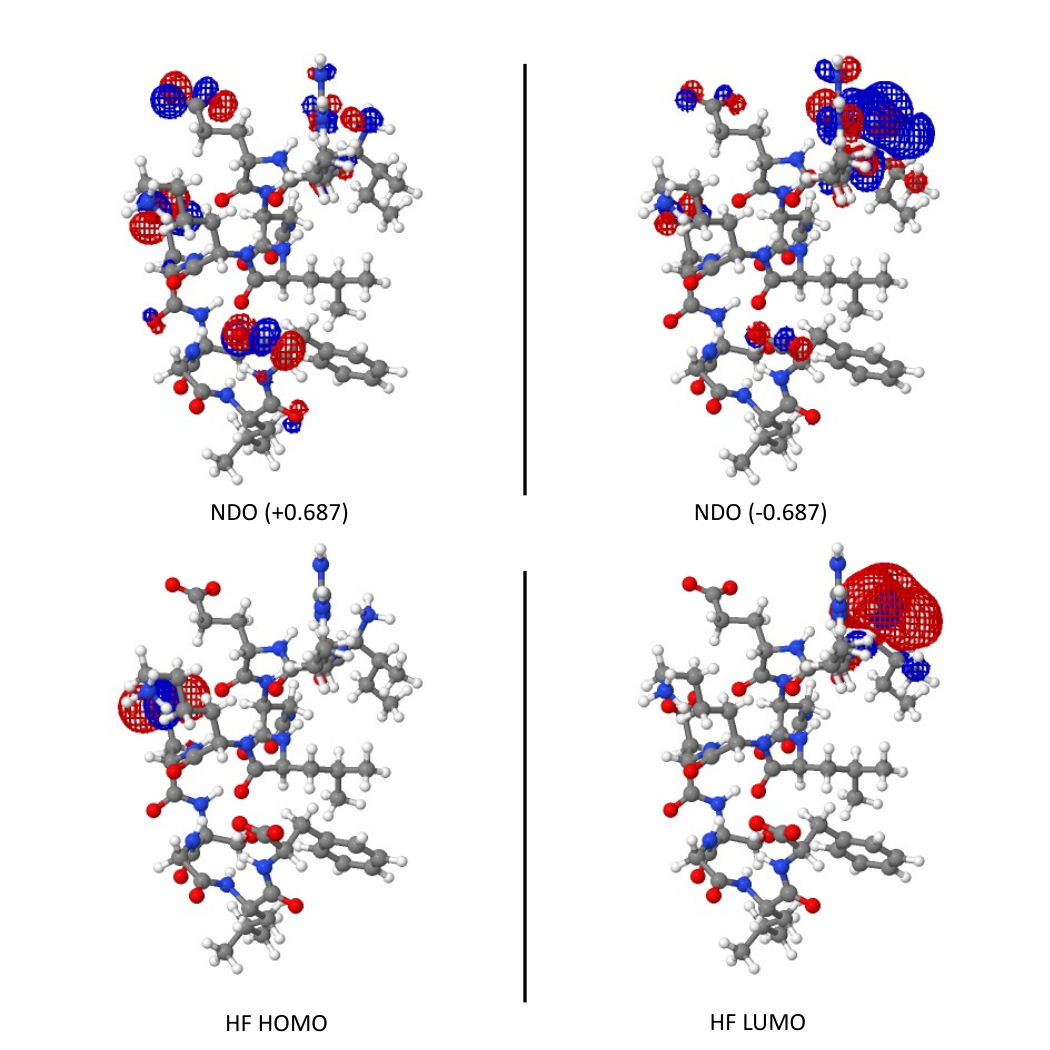}
        \caption{1LVZ: HF-LDA FNDOs, juxtaposed with the Hartree-Fock HOMO and LUMO.}
        \label{fig:1LVZ_nd_vs_hf}
        \end{figure}

        \begin{figure}
        \centering
        \includegraphics[width=7.0in]{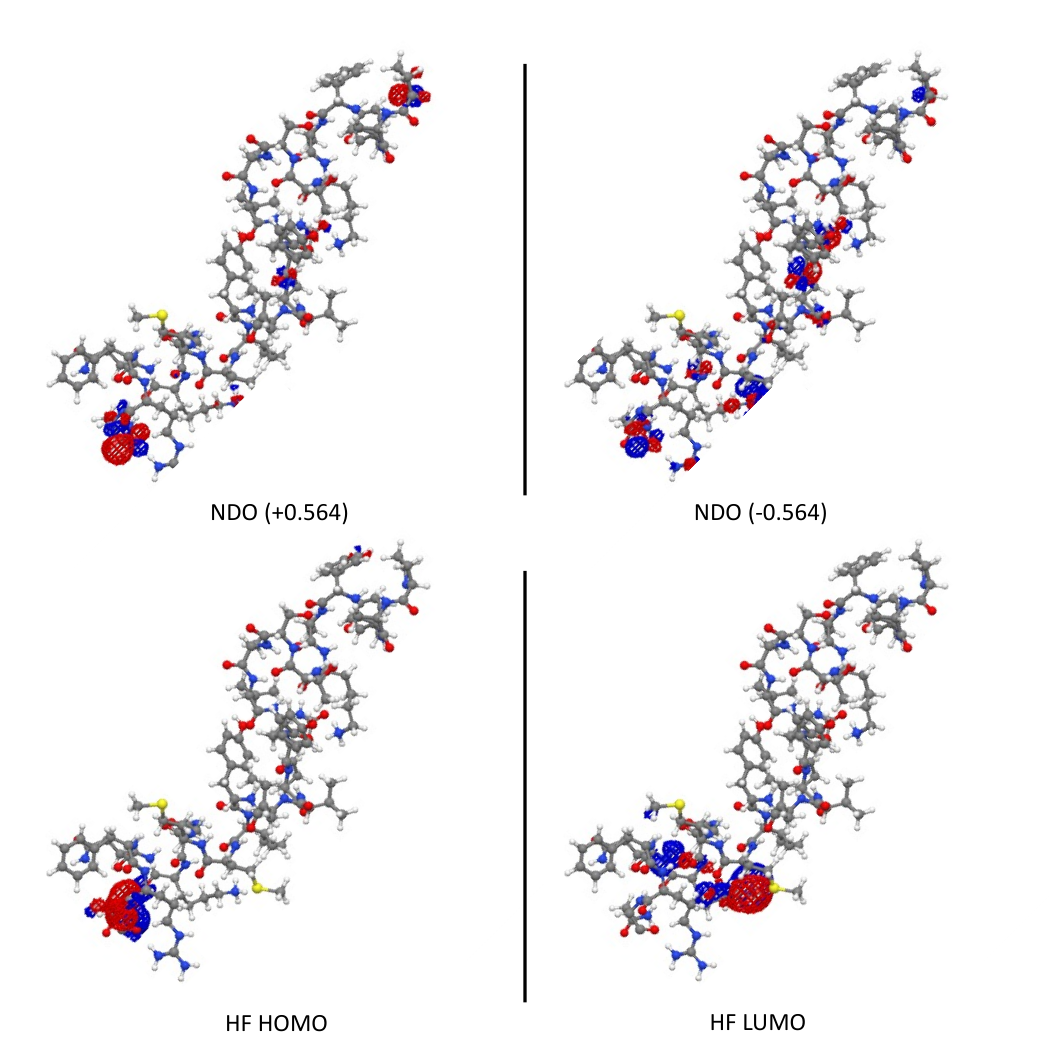}
        \caption{1FDF: HF-LDA FNDOs, juxtaposed with the Hartree-Fock HOMO and LUMO.}
        \label{fig:1FDF_nd_vs_hf}
        \end{figure}

        \begin{figure}
        \centering
        \includegraphics[width=7.0in]{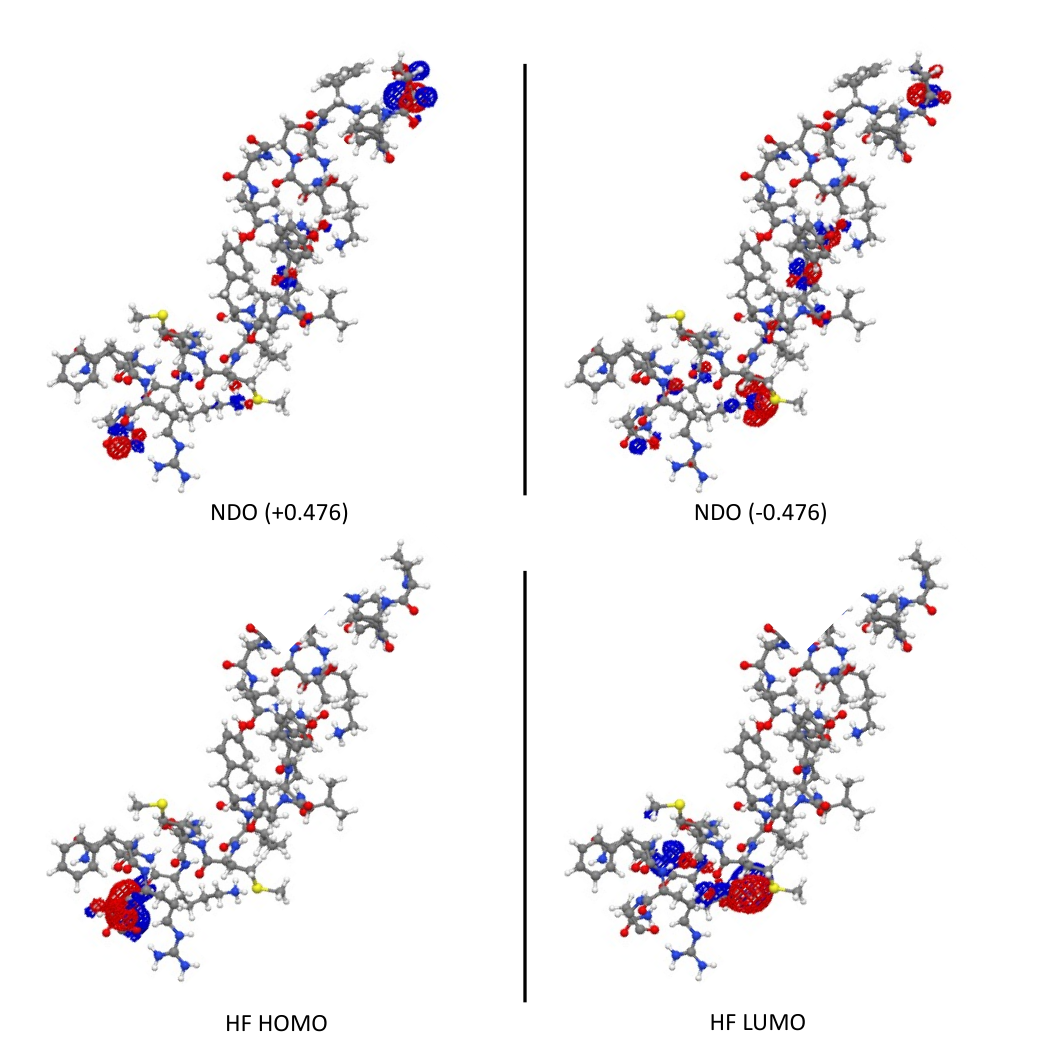}
        \caption{1FDF: HF-LDA FNDOs, juxtaposed with the Hartree-Fock HOMO and LUMO.}
        \label{fig:1FDF_nd_vs_hf_2nd}
        \end{figure}